\documentclass[12pt,epsf]{article}
\usepackage{amsmath}
\usepackage{amssymb}
\begin{document}
\def\a{\alpha}
\def\b{\beta}
\def\c{\varepsilon}
\def\d{\delta}
\def\e{\epsilon}
\def\f{\phi}
\def\g{\gamma}
\def\h{\theta}
\def\k{\kappa}
\def\l{\lambda}
\def\m{\mu}
\def\n{\nu}
\def\p{\psi}
\def\q{\partial}
\def\r{\rho}
\def\s{\sigma}
\def\t{\tau}
\def\u{\upsilon}
\def\v{\varphi}
\def\w{\omega}
\def\x{\xi}
\def\y{\eta}
\def\z{\zeta}
\def\D{{\mit \Delta}}
\def\G{\Gamma}
\def\H{\Theta}
\def\L{\Lambda}
\def\F{\Phi}
\def\P{\Psi}

\def\S{\Sigma}

\def\o{\over}
\def\beq{\begin{eqnarray}}
\def\eeq{\end{eqnarray}}
\newcommand{\gsim}{ \mathop{}_{\textstyle \sim}^{\textstyle >} }
\newcommand{\lsim}{ \mathop{}_{\textstyle \sim}^{\textstyle <} }
\newcommand{\vev}[1]{ \left\langle {#1} \right\rangle }
\newcommand{\bra}[1]{ \langle {#1} | }
\newcommand{\ket}[1]{ | {#1} \rangle }
\newcommand{\EV}{ {\rm eV} }
\newcommand{\KEV}{ {\rm keV} }
\newcommand{\MEV}{ {\rm MeV} }
\newcommand{\GEV}{ {\rm GeV} }
\newcommand{\TEV}{ {\rm TeV} }
\def\slash#1{\ooalign{\hfil/\hfil\crcr$#1$}}
\def\diag{\mathop{\rm diag}\nolimits}
\def\Spin{\mathop{\rm Spin}}
\def\SO{\mathop{\rm SO}}
\def\O{\mathop{\rm O}}
\def\SU{\mathop{\rm SU}}
\def\U{\mathop{\rm U}}
\def\Sp{\mathop{\rm Sp}}
\def\SL{\mathop{\rm SL}}
\def\tr{\mathop{\rm tr}}
\baselineskip 0.7cm

\begin{titlepage}

\begin{flushright}
UCB-PTH-08/70
\end{flushright}

\vskip 1.35cm
\begin{center}
{\large \bf Non-compact Mirror Bundles and $(0,2)$ Liouville Theories}

\vskip 2.2cm

Yu Nakayama

\vskip 0.4cm
{\it Berkeley Center for Theoretical Physics and Department of Physics}
\\

\vskip 3.5cm

\abstract{We study $(0,2)$ deformations of $\mathcal{N}=2$ Liouville field theory and its mirror duality. A gauged linear sigma model construction of the ultraviolet theory connects $(0,2)$ deformations of Liouville field theory and $(0,2)$ deformations of $\mathcal{N}=2$ $SL(2,\mathbf{R})/U(1)$ coset model as a mirror duality.  Our duality proposal from the gauged linear sigma model completely agrees with the exact CFT analysis. In the context of heterotic string compactifications, the deformation corresponds to the introduction of a non-trivial gauge bundle. This non-compact Landau-Ginzburg construction yields a novel way to study the gauge bundle moduli for non-compact Calabi-Yau manifolds.}
\end{center}
\end{titlepage}

\setcounter{page}{2}

\section{Introduction}
$(0,2)$ superconformal field theories (SCFTs) in two dimension occupy a huge landscape of perturbative heterotic string theories with $\mathcal{N}=1$ supersymmetric compactifications, yet a significant portion of the landscape is not fully scrutinized and remains to be investigated. At the classical level, the $(0,2)$ deformations of the heterotic non-linear sigma models correspond to bundle deformations of heterotic $E_8 \times E_8$ gauge group away from the so-called standard embedding. However, it is generically expected that the world-sheet instanton corrections in the $(0,2)$ non-linear sigma model break the conformal invariance and the background does not make sense as a string world-sheet theory.

In other words, the world-sheet instanton effects will give rise to a non-perturbative superpotential for such ``would-be moduli" from the target space-time viewpoint.\footnote{Under certain conditions \cite{Distler:1987ee}, the instanton corrections vanish, sometimes due to seemingly miraculous cancellation\cite{Beasley:2003fx}. In this paper, we focus on the gauged linear sigma model (GLSM) construction, where it is known that the world-sheet instanton effects do not break the conformal invariance \cite{Silverstein:1995re}\cite{Basu:2003bq}.} The resulting generation of the non-perturbative superpotential would play a fundamental role in investigating the moduli stabilization problem in heterotic compactifications. In this sense, the study of the heterotic string theory focusing only on $(2,2)$ locus merely scratches the whole surface of the heterotic landscape: the huge $(0,2)$ space remains un-explored.

From the world-sheet viewpoint, the $(2,2)$ locus is much easier to study. For example, the celebrated mirror symmetry \cite{Hori:2003ic} (as generalized T-duality) plays a fundamental role in understanding the topological (or BPS) nature of the $(2,2)$ string compactification, and it has provided us a fruitful interconnection between mathematics and the string theory.

A similar situation may be expected in $(0,2)$ SCFTs. Indeed, there have been mathematical as well as physical approaches to generalize the concept of mirror symmetry to $(0,2)$ SCFTs \cite{Blumenhagen:1996vu}\cite{Blumenhagen:1996tv}\cite{Blumenhagen:1997pp}\cite{Knutson:1997yt}\cite{Sharpe:1998wa}\cite{Adams:2003zy}\cite{Guffin:2007mp}\cite{Melnikov:2007xi}\cite{Guffin:2008pi}\cite{McOrist:2008ji}. In this paper, we would like to further investigate this $(0,2)$ mirror symmetry from the generalization of the duality between the  sine-Liouville theory and the 2D-black hole, which is known as Fateev-Zamolodchikov-Zamolodchikov (FZZ) duality \cite{FZZ}.\footnote{See \cite{Nakayama:2004vk} for a review about the Liouville field theory and related topics.}

One of the key features of the string theory is that it enables us to understand the resolution of singularity from different perspectives. The FZZ duality is such an example, relating the (winding) tachyon condensation and the geometric resolution of singularity as a duality between non-compact Calabi-Yau space and non-compact Gepner models. It has been noticed \cite{Hori:2001ax} that indeed the $(2,2)$ version of the FZZ duality may be understood as a mirror symmetry, and this idea has led to the proof of the duality. 

In this paper, we study the $(0,2)$ version of the FZZ duality. Here, again, the world-sheet non-perturbative effects play an important role, and a possible structure of the conformal gauge bundle deformations crucially depends on the form of the instanton corrections.
We will further give a geometric interpretation of the duality as deformations of the vector bundle moduli in non-compact Calabi-Yau space. The non-compact Calabi-Yau space is a clean setup to study the localized structure of the string theory, where the gravity is decoupled from the localized degrees of freedom. We hope that our study will become a first step to understand the local mirror symmetry of the non-compact Calabi-Yau space with non-trivial gauge bundle deformations. 

The organization of the rest of the paper is as follows. In section 2, we briefly review $(0,2)$ superspace and superfields to establish our notation. In section 3, we provide some basic aspects of $(0,2)$ mirror symmetry from the world-sheet viewpoint. In section 4, we construct an example of $(0,2)$ non-compact mirror symmetry as a $(0,2)$ version of the FZZ duality, which turns out to be an irrelevant deformation. In section 5, we show an example with the non-trivial vector bundle deformation even as a CFT. In section 6, we interpret our world-sheet results from the space-time geometric viewpoint. In section 7, we conclude our paper with some discussions. We dedicated two appendices to review relevant aspects of $SL(2,\mathbf{R})/U(1)$ Kazama-Suzuki coset model and $\mathcal{N}=2$ Liouville theory.

\section{$(0,2)$ superspace, $(0,2)$ superfield}
In this section, we establish our convention for $(0,2)$ supersymmetry in two-dimension with $(0,2)$ superspace and $(0,2)$ superfields. Some useful references are \cite{Witten:1993yc}\cite{Adams:2003zy}.

$(0,2)$ supersymmetry in two-dimension\footnote{Our convention is that left-moving  $=$ holomorphic: $f(x-t) = f(z)$, and right-moving $=$ anti-holomorphic: $g(x+t) = g(\bar{z})$. In heterotic string theory, the right-mover is supersymmetric in our convention.} is generated by two fermionic supercharges $Q_+$, $\bar{Q}_+ = Q_+^\dagger$ together with bosonic generators: Hamiltonian $H$, momentum $P$, and rotation $M$ and (possibly\footnote{In $(0,2)$ superconformal theories, the $U(1)$ R-symmetry is necessary.}) $U(1)$ R-symmetry $F_+$. The commutation relation is 
\begin{eqnarray}
Q^2_+ = \bar{Q}_+^2 = 0 \ ,  \ \  \{Q_+, \bar{Q}_+\} = 2 (H-P) \ , \cr
[M,Q_+] = - Q_+ \ , \ \ [M,\bar{Q}_+] = - \bar{Q}_+ \ , \cr
[F_+,Q_+] = - Q_+ \ , \ \ [F_+,\bar{Q}_+] = + \bar{Q}_+ 
\end{eqnarray}
with obvious commutation relations for Poincare symmetry.

It is useful to use the $(0,2)$ superspace $(y^+,y^-, \theta^+,\bar{\theta}^+)$ to construct supersymmetric Lagrangian. The superderivatives are defined as
\begin{eqnarray}
D_+ = \frac{\partial}{\partial \theta^+} - i\bar{\theta}^+ \partial_+  \ , \ \ \bar{D}_+ = -\frac{\partial}{\partial\bar{\theta}^+} + i \theta^+ \partial_+ \ ,
\end{eqnarray}
which satisfy
\begin{eqnarray}
\{D_+,D_+\} = \{\bar{D}_+,\bar{D}_+\} = 0 \ , \ \{\bar{D}_+,D_+\} = 2i \partial_+ \ .
\end{eqnarray}

\subsection{Chiral Multiplet}
A chiral superfield is defined by the condition $\bar{D}_+ \Phi = 0$. In the component form, it contains a complex scalar $\phi(y)$ and a complex Weyl fermion $\psi(y)$ as
\begin{eqnarray}
\Phi = \phi(y) + \sqrt{2}\theta^+ \psi_+(y) - i \theta^+ \bar{\theta}^+ \partial_+ \phi (y) \ .
\end{eqnarray}
The free action is given by\footnote{Our convention is $\int d^2\theta = \frac{\partial}{\partial \bar{\theta}^+} \frac{\partial}{\partial \theta^+}$.}
\begin{align}
S &= -\frac{i}{2} \int d^2y d^2\theta \bar{\Phi} \partial_- \Phi \cr
&= \int d^2 y\left( -\partial_\mu \bar{\phi} \partial^\mu \phi + i \bar{\psi}_+ \partial_- \psi_+\right)
\end{align}
where $\bar{\Phi}$ is an anti-chiral superfield (complex conjugate of $\Phi$) satisfying $D_+ \bar{\Phi} = 0$.

\subsection{Fermi multiplet}
Fermi multiplet satisfies the condition 
\begin{eqnarray}
\bar{D}_+ \Gamma = \sqrt{2}E \ ,
\end{eqnarray}
where $E$ satisfies
\begin{eqnarray}
\bar{D}_+ E = 0 \ .
\end{eqnarray}
The component expansion of the fermi multiplet $\Gamma$ is given by
\begin{eqnarray}
\Gamma = \psi_- - \sqrt{2}\theta^+ G - i \theta^+ \bar{\theta}^+ \partial_+ \psi_- - \sqrt{2}\bar{\theta}^+ E \ ,
\end{eqnarray}
where $\psi_-$ is a complex Weyl fermion and $G$ is an auxiliary field.
The simple action with a conventional kinetic term is given by
\begin{align}
S &= -\frac{1}{2}\int d^2 y d^2\theta \bar{\Gamma}\Gamma \cr
&= \int d^2y \left(i\bar{\psi}_-\partial_+ \psi_- + |G|^2 - |E|^2 - \bar{\psi}_- \frac{\partial E}{\partial \phi_i} \psi_{+i} - \bar{\psi}_{+i} \frac{\partial \bar{E}}{\partial \bar{\phi}_i} \psi_- \right) \ , 
\end{align}
where we have assumed that $E$ is a holomorphic function of chiral superfields $\Phi_i$. $(2,2)$ chiral multiplet is decomposed into one $(0,2)$ chiral multiplet and one fermi multiplet.

Furthermore, we can add superpotential terms. By definition, it is given by an integration over half the superspace:
\begin{align}
S &= -\frac{1}{\sqrt{2}} \int d^2y d\theta^+ \Gamma_a J^a|_{\bar{\theta}^+=0} - h.c. \cr
&= - \int d^2y \left(G_aJ^a + \psi_{-a}\psi_{+i} \frac{\partial J^a}{\partial \phi_i} \right)  + h.c.  \ 
\end{align}
with a holomorphic function $J^a(\Phi)$, 
where the $(0,2)$ supersymmetry requires $E_aJ^a =0$. When $E=0$ and $J^a = \partial^a W$ (in addition to the canonical kinetic term as above), we have an enhanced $(2,2)$ supersymmetry.

\subsection{Vector multiplet}
Next we study $U(1)$ gauge multiplet.
We define covariant superderivatives by
$\mathcal{D}_+ = e^{-\Psi} D_+ e^{\Psi}$ and $\bar{\mathcal{D}}_+ = e^{\bar{\Psi}} \bar{D}_+ e^{-\bar{\Psi}}$. The connection $\Psi$ has a gauge transformation
\begin{align}
\delta \Psi &= i \Lambda \  \cr
\delta\bar{\Psi} &= - i \bar{\Lambda}  ,
\end{align}
where $\Lambda$ is a chiral superfield (i.e. $\bar{D}_+ \Lambda = 0$). We can easily see
\begin{eqnarray}
\mathcal{D}^2_+ = \bar{\mathcal{D}}^2_+ = 0 \ ,  \ \  \{ \mathcal{D}_+, \bar{\mathcal{D}}_+\} = 2i (\mathcal{D}_0 + \mathcal{D}_1) \ .
\end{eqnarray}
We can use the gauge invariance to impose the Wess-Zumino gauge condition\footnote{As we will see, only the combination $\Psi + \bar{\Psi}$ appears in the action.}
\begin{eqnarray}
\Psi + \bar{\Psi} = \theta^+ \bar{\theta}^+ (A_+) \ ,
\end{eqnarray}
which is equivalent to 
\begin{align}
\mathcal{D}_0 + \mathcal{D}_1 &= \partial_+ - D_+\bar{D}_+ (\Psi+\bar{\Psi}) = \partial_+ + i A_+ \cr
\mathcal{D}_+ &= \frac{\partial}{\partial \theta^+} - i \bar{\theta}^+(\mathcal{D}_0 + \mathcal{D}_1) \cr
\bar{\mathcal{D}}_+  &= -\frac{\partial}{\partial \bar{\theta}^+} + i \theta^+(\mathcal{D}_0 + \mathcal{D}_1) \ ,
\end{align}
where $A_+$ is the right-moving connection.

The left-moving connection is independent of $\Psi$ and defined by a (real valued) vector superfield $V$ as
\begin{eqnarray}
V = A_- - \sqrt{2}i \theta^+ \bar{\lambda}_- - \sqrt{2}i \bar{\theta}^+ \lambda_- + 2\theta^+ \bar{\theta}^+ D \ 
\end{eqnarray}
so that
\begin{eqnarray}
\mathcal{D}_0 - \mathcal{D}_1 = \partial_- + i V \ ,
\end{eqnarray}
where $A_-$ is the left-moving connection and $\lambda_-$ is the left-moving gaugino while $D$ is an auxiliary field.
The connection $V$ has the gauge transformation
\begin{eqnarray}
\delta V = \partial_- (\Lambda + \bar{\Lambda}) \ .
\end{eqnarray}

The gauge invariant field strength is defined by
\begin{align}
\Upsilon &= \bar{D}_+ (\partial_-({\Psi}+\bar{\Psi}) + iV)  \cr
&=  -(\sqrt{2}\lambda_- -i 2\theta^+(D-iF_{01}) - \sqrt{2}i\theta^+\bar{\theta}^+ \partial_+ \lambda_-) \ .
\end{align}
The conventional kinetic term is given by
\begin{align}
S &= -\frac{1}{8e^2} \int d^2y d^2 \theta \bar{\Upsilon} \Upsilon \cr
&= \frac{1}{2e^2} \int d^2 y \left(F_{01}^2 + i \bar{\lambda}_- \partial_+\lambda_- + D^2 \right) \ . 
\end{align}
One can also introduce the FI term
\begin{align}
S_{FI} &= \frac{t}{4} \int d^2y d\theta^+ \Upsilon|_{\bar{\theta}^+ = 0} + h.c. \cr
&=  \frac{t}{2}\int d^2y (F_{01} + iD) + h.c. \ .
\end{align}

Finally, the right handed gaugino is not in the gauge multiplet but belongs to a chiral multiplet as
\begin{eqnarray}
\Sigma = \sigma + \sqrt{2} \theta^+{\lambda}_+ - i \theta^+\bar{\theta}^+ \partial_+ \sigma \ .
\end{eqnarray}
$(2,2)$ supersymmetry demands $E_a = Q \Sigma \Phi_a$ with the conventional kinetic term for the right-handed gaugino
\begin{align}
S &= -\frac{i}{4e^2} \int d^2y d^2\theta \bar{\Sigma}\partial_- \Sigma \cr
  &= \int d^2 y \frac{1}{2e^2} (-\partial^\mu \bar{\sigma}\partial_\mu \sigma + i \bar{\lambda}_+\partial_- \lambda_+) \ .
\end{align}

\subsection{Gauge invariant matter}
A chiral superfield (i.e. $\bar{D}_+ \Phi =0$) with charge $Q$  transforms as $\delta \Phi = e^{2iQ\Lambda} \Phi$ under the gauge transformation. The invariant action should be
\begin{align}
S &= -\frac{i}{2} \int d^2y d^2 \theta \frac{1}{2}\left(\bar{\Phi} e^{-2Q\bar{\Psi}} (2 \partial_- + 2Q\partial(\Psi-\bar{\Psi})- iV2Q) e^{-2Q\Psi} \Phi \right) \cr
&= -\frac{i}{2} \int d^2y d^2 \theta \frac{1}{2}\left(\bar{\Phi}e^{-2Q(\Psi+ \bar{\Psi})}\partial_-\Phi - \Phi e^{-2Q(\Psi + \bar{\Psi})} \partial_- \bar{\Phi} - 2iQ \bar{\Phi} e^{-2Q(\Psi +\bar{\Psi})} V \Phi\right) \cr
&= \int d^2 y  \left( -|D_\mu\phi|^2 + i \bar{\psi}_+(D_0-D_1)\psi_+ - iQ\bar{\phi} \lambda_- \psi_+ + iQ\phi\bar{\psi}_+\bar{\lambda}_- + QD|\phi|^2 \right) \ , \label{chirals}
\end{align}
where the covariant derivative is $D_\mu \phi = (\partial_\mu -iQA_\mu)\phi$.

Similarly for fermi superfield with charge $Q$ (i.e. $\delta \Gamma = e^{2iQ \Lambda} \Gamma$), we have the gauge invariant action
\begin{align}
S &= -\frac{1}{2}\int d^2y d^2\theta \bar{\Gamma} e^{-2Q(\Psi + \bar{\Psi})} \Gamma \cr
&= \int d^2 y \left( i \bar{\chi}_-(D_0 + D_1) \chi_- + |G|^2 - |E|^2 - \bar{\chi}_- \frac{\partial E}{\partial \phi_i} \psi_{+i} - \bar{\psi}_{+i} \frac{\partial \bar{E}}{\partial \bar{\phi}_i} \chi_- \right)
\ . \end{align}

For later purposes, we also study the axionic (shift) gauge symmetry, $\delta P = 2iQ\Lambda$. The invariant action is
\begin{align}
 S &=-\frac{i}{2} \int d^2y d^2 \theta \frac{1}{2}(P + \bar{P} -2Q(\Psi + \bar{\Psi}))(\partial_- {P} -\partial_- \bar{P}-i2QV) \cr
&= \int d^2 y \left(- |D_\mu p|^2+ i\bar{\chi}_+\partial_-\chi_+ + QD(p+\bar{p}) + Qi\chi_+ \lambda_- + Qi\bar{\chi}_+ \bar{\lambda}_- \right) \ ,
\end{align}
where $D_\mu p =\partial_\mu p -iQA_\mu$.

\section{Mirror Duality}
From the world-sheet theory viewpoint, the mirror symmetry can be understood as an S-duality of the GLSM whose infrared limit corresponds to the non-linear sigma model of the (mirror) geometry \cite{Hori:2000kt}\cite{Adams:2003zy}. In this section, we review the Abelian S-duality of the GLSM and summarize the general aspects of the $(0,2)$ mirror symmetry from the world-sheet perspective.

\subsection{Perturbative duality for chiral multiplet}
The idea to show Abelian S-duality is to transform the action into two different but equivalent forms by changing the order of Gaussian integration of quadratic fields. 
We begin with the following action
\begin{eqnarray}
\int d^2y d^2\theta\left[ -\frac{i}{4} e^{2B -2Q(\Psi + \bar{\Psi})} (-2QiV + 2i A) - iF \bar{D}_+(\partial_- B + iA) + iFD_+(\partial_- B - iA)\right] \ ,
\end{eqnarray}
where $A$ and $B$ are unconstrained real superfield, and $F$ is an unconstrained Lagrange multiplier fermi superfield. If we first integrate out $F$, we obtain
\begin{eqnarray}
2 B = \pi + \bar{\pi}  \ \ \ 2iA = \partial_-(\pi - \bar{\pi}) \ ,
\end{eqnarray}
where $\pi$ is a chiral superfield. Then, after substituting back into the original action, it becomes
\begin{eqnarray}
\int d^2 y d^2\theta -\frac{i}{4} e^{\pi + \bar{\pi} - 2Q(\Psi + \bar{\Psi})} [\partial_- (\pi - \bar{\pi}) - i2QV] \ .
\end{eqnarray} 
Introducing $\Phi = e^{\pi}$, we obtain the gauged action for a chiral multiplet in \eqref{chirals}. 

On the other hand, we can first integrate out $A$ and $B$ by introducing a chiral field $\frac{1}{4} Y = \bar{D}_+ F$, which gives
\begin{eqnarray}
2 B = + 2Q(\Psi + \bar{\Psi}) + \log \frac{Y + \bar{Y}}{2} \ \ \ 2iA = i2QV - \frac{\partial_-(Y-\bar{Y})}{Y+\bar{Y}} \ . 
\end{eqnarray}
Inserting this into the action, we obtain
\begin{eqnarray}
S_{\mathrm{dual}} = \frac{i}{8} \int d^2 y d^2\theta \frac{(Y-\bar{Y})\partial_-(Y+\bar{Y})}{(Y+\bar{Y})} - \frac{i}{4} \int d^2 y d\theta^+ QY \Upsilon + h.c. \ .
\end{eqnarray}

\subsection{Dual for axion superfield}
We begin with
\begin{eqnarray}
\int d^2y d^2\theta -\frac{i}{4} (2B -2Q(\Psi + \bar{\Psi})) (-2QiV + 2i A) - iF \bar{D}_+(\partial_- B + iA) + iF{D}_+(\partial_- B - iA) \ ,
\end{eqnarray}
Integrating out $F$ gives 
\begin{eqnarray}
2 B = \pi + \bar{\pi}  \ \ \ 2iA = \partial_-(\pi - \bar{\pi}) \ ,
\end{eqnarray}
which result in
\begin{eqnarray}
\int d^2 y d^2\theta -\frac{i}{4} (\pi + \bar{\pi}-2 Q(\Psi + \bar{\Psi})) [\partial_- (\pi - \bar{\pi}) - i2QV] \ .
\end{eqnarray} 
This is the action for the axionic chiral multiplet.

On the other hand, if we first integrate out $A$ and $B$ with introducing a chiral superfield $Y_P$ as $\frac{1}{2}Y_P = \bar{D}_+F$, we have
\begin{eqnarray}
2 B = 2Q(\Psi + \bar{\Psi}) + Y_P + \bar{Y}_P \  \ \ 2iA = i2QV - \partial_-(Y_P-\bar{Y}_P) \ .
\end{eqnarray}
Substituting back into the original action, we obtain
\begin{eqnarray}
S_{\mathrm{dual}} = \frac{i}{4} \int d^2 yd^2\theta (Y_P-\bar{Y}_P)\partial_-(Y_P+\bar{Y}_P) -\frac{i}{2}\int d^2y d\theta^+ QY_P\Upsilon + h.c. \ .
\end{eqnarray}

\subsection{Dual for fermi multiplet}
The starting point is 
\begin{eqnarray}
\int d^2 y d^2 \theta -\frac{1}{2}\bar{N} e^{-2Q(\Psi+\bar{\Psi})} N + S(\bar{D}_+ N - \sqrt{2}E) -\bar{S} (D_+ \bar{N}+\sqrt{2}E) \ 
\end{eqnarray}
with an unconstrained superfield $S$.
Integrating out $S$ first gives 
\begin{eqnarray}
\bar{D}_+ N = \sqrt{2} E \ ,
\end{eqnarray}
which is solved by $N = \Gamma$. Then the action becomes
\begin{eqnarray}
- \frac{1}{2} \int d^2 y d^2 \theta \bar{\Gamma} e^{-2Q(\Psi+\bar{\Psi})} \Gamma \ .
\end{eqnarray}

On the other hand, if one solves $N$ first, then
\begin{eqnarray}
 \bar{D}_+ S = -\frac{1}{2} \bar{N} e^{-2Q(\Psi + \bar{\Psi})} \ .
\end{eqnarray}
One can define a chiral superfield $G = \bar{N}e^{-2Q(\Psi + \bar{\Psi})}$ so that $\bar{D}_+ G = 0$, and the action becomes
\begin{eqnarray}
S_{\mathrm{dual}} = \int d^2 y d^2\theta -\frac{1}{2}\bar{G}e^{2Q(\Psi+\bar{\Psi})}G -\left( \int d^2y d\theta^+ \frac{1}{\sqrt{2}} G E  + h.c. \right) \ .
\end{eqnarray}
If one wants to dualize $E(\Phi)$ at the same time, we can first introduce a neutral superfield $F=GE$ and the action becomes
\begin{eqnarray}
S_{\mathrm{dual}} = -\frac{1}{2}\int d^2y d^2 \theta \frac{\bar{F}F}{\bar{E}e^{-2Q(\Psi+\bar{\Psi})} E} - \left( \int d^2y d\theta^+ \frac{1}{\sqrt{2}} F + h.c. \right)\ ,
\end{eqnarray}
where we have to express $\bar{E}e^{-2Q(\Psi+\bar{\Psi})} E$ in terms of the dual variable $Y$.

\subsection{Non-perturbative superpotential}
There exist additional contributions to the superpotential coming from non-perturbative instanton (vortex) effects. For simplicity, and sufficiently for our purposes, we assume $E_a = \Sigma a_{ai}\Phi_i$ for chiral multiplets or $E_P = \Sigma$ for an axionic multiplet. The theory has a vector R-symmetry (i.e. $Q(\theta) = -1$): $Q(\Phi^i) = Q(P) = Q(\Sigma)=0$, $Q(\Gamma ) = Q(\Gamma_P) = 1$ and $Q(\Upsilon) = -1$. In the component form, we have\footnote{Throughout this paper, we stick to the convention $\psi_{\pm}$ denotes the fermion for linearly charged chiral multiplet $\Phi_i$ (and $\Gamma_i$), while $\chi_{\pm}$ denotes the fermion for axionic chiral multiplet $P$ (and $\Gamma_P$).} 
\begin{eqnarray}
\psi_\pm \to e^{i\alpha} \psi_{\pm} \cr
\chi_\pm \to e^{i\alpha} \chi_{\pm} \cr
\lambda_{\pm} \to e^{-i\alpha} \lambda_{\pm}
\end{eqnarray}

 There is also an axial R-symmetry: $Q(\Phi) = Q(P)=0$, $Q(\Gamma) = Q(\Gamma_P) = Q(\Upsilon) =-1$, and $Q(\Sigma) = -2$. In the component form, we have
\begin{eqnarray}
\psi_{\pm} \to e^{\pm i\alpha} \psi_{\pm} \cr
\chi_{\pm} \to e^{\pm i\alpha} \chi_{\pm} \cr
\sigma  \to e^{-2i\alpha} \sigma \cr
\lambda_{\pm} \to e^{\pm i\alpha} \lambda_{\pm} \cr .
\end{eqnarray}
Without the axionic multiplet, the axial current is anomalous, but one could improve the axial current by adding the gauge invariant quantity $\mathcal{A}_\mu = \partial_\mu (\mathrm{Im} P) + Q_PA_\mu$ to cancel the anomaly. 

The examination of the possible instanton configuration and the symmetry discussion above constrain the form of the non-perturbative superpotential in the dual theory severely (but not completely) \cite{Hori:2001ax}\cite{Adams:2003zy}. Under the axial symmetry, the dual field is shifted as $Y_i \to Y_i -i \alpha$ to realize the anomaly. The anomaly cancellation demands that $Y_P$ transforms as $Y_P \to Y_P + i \frac{\sum_i Q_i}{2Q_P} \alpha$.

From the BPS nature of the superpotential and the above symmetry argument, the non-perturbative superpotential takes the following general form: 
\begin{eqnarray}
W_{\mathrm{non-pert}} =  (\alpha_{ia} F_i + \beta_{a}F_{P})  e^{-Y_a} + (\tilde{\alpha}_{i} F_i + \tilde{\beta}F_{P}) e^{\frac{2Q_P}{\sum_i{Q_i}} Y_{P}} \ .
\end{eqnarray}
In $(2,2)$ limit, $\beta_{a} = \tilde{\alpha}_{i} = \tilde{\beta}=0 $ \cite{Hori:2001ax}. It seems possible to repeat the argument given in \cite{Hori:2001ax} to conclude this is also true for $(0,2)$ theory. First, we split the gauge symmetry for $\Phi_i$ and $P$ by $U(1)$ and $U(1)_P$. Then, there is no non-perturbative corrections for $Y_P$ because it is just a free massive vector theory. Now, one can freeze $\Sigma - \Sigma_P$ and $\Upsilon - \Upsilon_P$ by tuning $D$-term couplings. The $D$-term cannot affect the superpotential term so this gives vanishing $\tilde{\alpha}_{i}$ and $\beta$.

This argument is not completely convincing, however, because we only have $(0,2)$ supersymmetry and $D$-term dependent non-perturbative correction might appear. Fortunately in our particular application with $E_a= a_{ai}\Sigma  \Phi_i$, since there is no vortex solution associated with $P$, one can argue that they must disappear exactly or at least can be absorbed by the redefinition of other superpotential coefficients. We first note that after the duality, gauge multiplets $\Sigma$ and $\Upsilon$ are all massive and can be integrated out, giving the condition
\begin{eqnarray}
Y_P = -\frac{\sum_i Y_i Q_i} {2Q_P} \  \ \ 
\end{eqnarray}
and a similar linear relation for $F_P$. Now, the effective superpotential for $Y_P$, if any, becomes
\begin{eqnarray}
\sim \sum_a c_{a} F_a e^{-\frac{\sum_i Y_i Q_i}{\sum_i Q_i}} \ .
\end{eqnarray}
If we have only one chiral field, then this term is the same as the non-perturbative superpotential for $Y$, so one can absorb it. If we have many chiral fields, then such a term cannot occur in the instanton computation because the BPS nature of the instanton computation forbids such a fractional contribution.

\section{Irrelevant $(0,2)$ deformation of $\mathcal{N}=2$ Liouville theory and its dual}
In this section, we propose a first example of $(0,2)$ Liouville duality. It turns out that the deformation is irrelevant in the far infrared regime.
\subsection{GLSM construction and low energy action}
A GLSM realizing $\mathcal{N}=2$ $SL(2,\mathrm{R})/U(1)$ supercoset model is given by one pair of a chiral multiplet $\Phi$ and a fermi multiplet $\Gamma$ together with one axion superfield $\Phi_P$ and its fermi partner $\Gamma_P$ \cite{Hori:2001ax}. The $(2,2)$ $U(1)$ gauge multiplet is realized by one pair of the vector multiplet $(V,\Psi)$ and a neutral chiral multiplet $\Sigma$. The non-trivial $(0,2)$ deformation we  will consider here is given by taking
\begin{align}
E &= \alpha Y \Sigma  \cr
E_Y &= \beta \Sigma  \ ,
\end{align}
where $\alpha = \beta = 1$ corresponds to the $(2,2)$ point. 

The superfield action is given by
\begin{align}
S & = -\frac{i}{2} \int d^2y d^2 \theta \frac{1}{2}\left(\bar{\Phi} e^{-2\bar{\Psi}} (2 \partial_- +2 \partial(\Psi-\bar{\Psi})- 2iV) e^{-2\Psi} \Phi \right)   -\frac{1}{2}\int d^2y d^2\theta \bar{\Gamma} e^{-2(\Psi + \bar{\Psi})} \Gamma \cr
 &  -\frac{i}{2} \int d^2y d^2 \theta \frac{k}{4}(P + \bar{P} -2(\Psi + \bar{\Psi}))(\partial_- \bar{P} -\partial_- P-i2V) -\frac{k}{4}\int d^2y d^2\theta \bar{\Gamma}_P \Gamma_P \cr
&-\frac{i}{4e^2} \int d^2 y d^2 \theta \bar{\Sigma} \partial_- \Sigma -\frac{1}{8e^2} \int d^2y d^2 \theta \bar{\Upsilon} \Upsilon \ .
\end{align}
The corresponding component Lagrangian is given by
\begin{align}
L &= - D^\mu \bar{\phi} D_{\mu} \phi + i \bar{\psi}_-(\partial_+ + i v_+)\psi_- + D|\phi|^2 +  i\bar{\psi}_+(\partial_- + iv_-) \psi_+ \cr
&- |\alpha|^2|\sigma|^2|\phi|^2 - \alpha \bar{\psi}_- \sigma \psi_+ -\bar{\alpha} \bar{\psi}_+ \bar{\sigma} \psi_- - i\bar{\phi} \lambda_- \psi_+ + i\bar{\alpha}  \bar{\phi} \lambda_+ \psi_- + i\bar{\psi}_+ \bar{\lambda}_- \phi - i \alpha \bar{\psi}_- \bar{\lambda}_+ \phi \cr
& + \frac{k}{2}\left(-(\partial_\mu p - iv_\mu)(\partial^\mu \bar{p} + iv^\mu) + i\bar{\chi}_-\partial_+ \chi_- + i \bar{\chi}_+\partial_- \chi_+ + D(p+\bar{p})  \right. \cr
& \left. -|\beta|^2|\sigma|^2 + i \chi_+ \lambda_- - i\bar{\beta} \chi_- \lambda_++ i \bar{\chi}_+\bar{\lambda}_- - i \beta \bar{\chi}_- \bar{\lambda}_+ \right) \cr 
& + \frac{1}{2e^2}(-\partial^\mu \bar{\sigma}\partial_\mu \sigma + i\bar{\lambda}_- \partial_+ \lambda_- + i \bar{\lambda}_+ \partial_- \lambda_+ + v_{01}^2 + D^2) \ . 
\end{align}

At low energy, one can first integrate out $\Sigma$ multiplet. Integrating out $\sigma$ gives four-fermi interaction
\begin{eqnarray}
-\frac{|\alpha|^2 \bar{\psi}_- \bar{\psi}_+ \psi_+ \psi_-}{\frac{k}{2}|\beta|^2+\alpha^2 |\phi|^2} \ .
\end{eqnarray}
Furthermore, integrating out gauginos $\lambda_\pm$ yields the relation
\begin{align}
\bar{\phi} \psi_+ &= -\frac{k}{2}\chi_+ \cr
\frac{\bar{\alpha}}{\bar{\beta}} \bar{\phi} \psi_- &= -\frac{k}{2}\chi_- \ . \label{ferr}
\end{align}

We can now choose the gauge $\mathrm{Im} p = 0$, and solve the $D$-term condition  as $ |\phi|^2 = - k \mathrm{Re} p$. The effective low energy dynamics for the remaining degrees of freedom is given by $(0,2)$ non-linear sigma model for ($\phi$,$\psi_{\pm}$). The bosonic part of the Lagrangian is given by
\begin{align}
L_B &= - \partial_\mu \bar{\phi} \partial^\mu \phi - \frac{1}{2k} \partial_\mu\phi \partial^\mu\phi \bar{\phi}^2 - \frac{1}{k} \partial_\mu\bar{\phi} \partial^\mu \phi |\phi|^2 - \frac{1}{2k} \partial_\mu \bar{\phi} \partial^\mu \bar{\phi} \phi^2 \cr
& -\frac{1}{4} \frac{\partial_\mu\phi\partial^\mu\phi \bar{\phi}^2}{\frac{k}{2} + |\phi|^2} - \frac{1}{4} \frac{\partial_\mu \bar{\phi} \partial^\mu \bar{\phi} \phi^2}{\frac{k}{2}+|\phi|^2} + \frac{1}{2}\frac{\partial_\mu \phi \partial^\mu \bar{\phi} |\phi|^2}{\frac{k}{2} + |\phi|^2} \ . \label{metn}
\end{align}
The bosonic part of the Lagrangian here does not depend on the deformation parameters $\alpha $ and $\beta$ and it describes the sigma model with the target-space metric\footnote{The metric \eqref{metn} does not look like Hermitian, but actually it is even Kahler by an appropriate choice of the coordinate as is clear from the fact that it is equivalent to the undeformed metric derived from $(2,2)$  supersymmetry, where the Kahler structure is automatic.}
\begin{eqnarray}
 ds^2 = (1+\frac{r^2}{k}) dr^2 + \frac{d\theta^2}{(1+\frac{r^2}{k})} \ , \label{metricn}
\end{eqnarray}
where we have introduced new coordinate $\phi = \frac{r}{\sqrt{2}} e^{i\theta}$ \cite{Hori:2001ax}.

The fermionic part of the Lagrangian, on the contrary, shows the effect of $(0,2)$ deformation. It is given by
\begin{align}
L_F  &= i\bar{\psi}_- \partial_+ \psi_- + i \bar{\psi}_+ \partial_- \psi_+ + \frac{ik}{2}\bar{\chi}_- \partial_+ \chi_- + \frac{ik}{2}\bar{\chi}_+ \partial_- \chi_+  \cr
& + \frac{\bar{\psi}_- \psi_-}{\frac{k}{2}+|\phi|^2} \left(\frac{i}{2}\phi \partial_+ \bar{\phi} - \frac{i}{2}\bar{\phi}\partial_+\phi \right) + \frac{\bar{\psi}_+\psi_+}{\frac{k}{2}+|\phi|^2} \left(\frac{i}{2}\phi\partial_-\bar{\phi}-\frac{i}{2}\bar{\phi}\partial_-\phi\right)   \cr
& - \frac{1}{\frac{k}{2} + |\phi|^2} (\bar{\psi}_-\psi_-\bar{\psi}_+\psi_+) - \frac{|\alpha|^2 \bar{\psi}_- \bar{\psi}_+ \psi_+ \psi_-}{ \frac{k}{2}|\beta|^2 +|\alpha|^2 |\phi|^2} \ ,
\end{align}
where we have to substitute \eqref{ferr} to remove $\chi_{\pm}$. Apart from the four-fermi term, the $(0,2)$ deformation comes only from this substitution, so the right-mover $\psi_+$ has the Riemann connection compatible with the metric \eqref{metricn} as is again clear from the fact that the right-moving part is not deformed from the $(2,2)$ locus. On the other hand, the left-moving fermion $\psi_-$ has a deformed connection corresponding to the non-trivial deformation of the gauge bundle away from the $(2,2)$ point: $V= TM$. After introducing the canonically normalized fermion as $\psi_- = \frac{\tilde{\psi}_-^{(0)}}{\sqrt{1+\frac{2}{k}\frac{\alpha^2}{\beta^2} |\phi|^2}}$, the first order perturbation in $\epsilon = 1-\frac{|\alpha|^2}{|\beta|^2}$ gives the deformation of the gauge bundle
\begin{eqnarray}
\delta A = -\frac{i}{k}\epsilon \frac{(\phi d \bar{\phi} - \bar{\phi} d \phi)}{1 + \frac{2}{k}|\phi|^2} \ .
\end{eqnarray}

The four-fermi interaction gives the field strength for the deformed gauge bundle $F$.
\begin{eqnarray}
\delta F = -2\frac{\frac{i}{k}\epsilon d\phi d\bar{\phi}}{1 + \frac{2}{k}|\phi|^2} +  \cdots
\end{eqnarray}
where ellipses represent higher order $O(1/k,\epsilon)$ corrections.

\subsection{Further renormalization}
The $(0,2)$ sigma model coupled with the non-trivial gauge bundle obtained in this way is classically a conformal field theory, but it is not quantum mechanically at the one-loop order. First of all, the metric \eqref{metricn} is not Ricci flat, so even when $\alpha = \beta$ (i.e. $(2,2)$ point), there is a non-trivial renormalization to make it conformal. It has been discussed in \cite{Hori:2001ax}, at $(2,2)$ point, the fixed point is given by $\mathcal{N}=2$ $SL(2,\mathbf{R})/U(1)$ Kazama-Suzuki coset model. The metric is given by the two-dimensional black hole \cite{2DBH}
\begin{eqnarray}
ds^2 = k(d\rho^2 + \tanh^2 \rho d\theta^2) \ ,
\end{eqnarray}
with the dilaton gradient
\begin{eqnarray}
\Phi = -2\log \cosh \rho \ .
\end{eqnarray}
In particular, the generation of the dilaton gradient is crucial to maintain the conformal invariance of the two-dimensional black hole background.

We here claim that the introduction of the $(0,2)$ deformation $\alpha \neq \beta$ is actually irrelevant for the IR physics, and the $(2,2)$ structure is recovered at the IR fixed point. Since the deformation is smoothly connected with the $(2,2)$ conformal fixed point, the non-zero deformation corresponds to an exactly marginal deformation of the $\mathcal{N}=2$ $SL(2,\mathbf{R})/U(1)$ coset model with preserving half amount of supersymmetry.

As we have reviewed in appendix (see also \cite{Hori:2001ax}), there is no such an exactly marginal deformation of the $\mathcal{N}=2$ $SL(2,\mathbf{R})/U(1)$ coset model \cite{Hori:2001ax} even if we relax the condition of non-zero momentum.\footnote{Here, we only focus on the deformation possible for any non-rational level $k$ because we are interested in the semiclassical deformation which is obtainable in the $k\to\infty$ limit of the non-linear sigma model. For specific values of $k$, there could be non-trivial deformation, which we would not discuss any further.} The only possible deformation descends from
\begin{align}
&J^{-}_{-1}\bar{J}^{-}_{-1}| j =1 \rangle^+ \otimes |0\rangle \cr 
&J^{+}_{-1}\bar{J}^{+}_{-1}| j =1 \rangle^- \otimes |0\rangle \ , \label{dudef}
\end{align}
or equivalently
\begin{align}
&[J^+_0 \tilde{J}_0^+|j=\frac{k}{2}\rangle^+ \otimes \psi_{-\frac{1}{2}}\bar{\tilde{\psi}}_{-\frac{1}{2}}|0\rangle ]^{w=-1} \cr
&[J^-_0 \tilde{J}_0^-|j=\frac{k}{2}\rangle^- \otimes \bar{\psi}_{-\frac{1}{2}}{\tilde{\psi}}_{-\frac{1}{2}}|0\rangle ]^{w=1} 
\end{align}
from the spectral flow isomorphism in the parent $SL(2,\mathbf{R})$ WZNW model. These are actually $(2,2)$ deformations (which are furthermore almost trivial, corresponding to renormalization of the $\mathcal{N}=2$ cosmological constant in the dual theory).

Thus, we conclude that there is no non-trivial exactly marginal $(0,2)$ deformation for the $\mathcal{N}=2$ $SL(2,\mathbf{R})/U(1)$ coset model, and the deformation introduced by $\alpha \neq \beta$ is irrelevant. This can be also seen from the dual Liouville description as we will see shortly.

\subsection{Dual theory}
The perturbative dual theory is obtained from the prescription reviewed in section 3. The kinetic term is given by
\begin{eqnarray}
L = \int d^2\theta\left[ \frac{i}{8}\frac{Y-\bar{Y}}{Y+\bar{Y}}\partial_-(Y + \bar{Y}) + \frac{i}{k} \bar{Y}_P \partial_- Y_P - \frac{\bar{F}F}{Y+\bar{Y}} - \frac{1}{k}\bar{F}_P F_P \right]\ ,
\end{eqnarray}
whereas the perturbative superpotential term is given by
\begin{eqnarray}
\int d\theta^+ \left(-\frac{i\Upsilon}{4}(Y+2Y_P) + \alpha \frac{\Sigma}{\sqrt{2}} F + \beta \frac{\Sigma}{\sqrt{2}} F_P \right) \ .
\end{eqnarray}
Furthermore, the non-perturbative superpotential term can be generated\footnote{As discussed before, the potentially allowed term $F_P e^{-\frac{1}{2}Y_P}$ can be absorbed by a redefinition of $\mu$.}
\begin{eqnarray}
\mu\int d\theta^+  F e^{-Y} \ 
\end{eqnarray}
from the instanton corrections.

In order to investigate the effective low energy action,
one can integrate out gauge multiplet to obtain the relation
\begin{eqnarray}
Y = -2Y_P  \  \ \  F = -\frac{\beta}{\alpha} F_P \ .
\end{eqnarray}
For a large real part of $Y$, the effective Lagrangian is given by
\begin{eqnarray}
 L = \int d^2 \theta \frac{i}{2}\frac{1}{2k}\bar{Y} \partial_- Y -\frac{|\alpha|^2}{|\beta|^2k}\bar{F} F-\left(\int d\theta^+ \mu F e^{-Y}  + h.c.\right)\ .
\end{eqnarray}
The leading order action preserves $(2,2)$ supersymmetry. This is broken by difference of the kinetic term between the bosonic field $Y$ and fermionic field $F$ for a smaller  real part of $Y$.

Our claim is that in the IR limit, the theory flows to $\mathcal{N}=2$ Liouville theory. First of all, we see that the superpotential term is not renormalized, and $(0,2)$ deformations only appear in the kinetic term. Since $\mathcal{N}=2$ Liouville theory, as in $\mathcal{N}=2$ $SL(2,\mathbf{R})/U(1)$ coset model, does not have any supersymmetric marginal deformation, the $(0,2)$ deformations should vanish at the conformal fixed point. In particular, the renormalization will generate a linear dilaton term $\int d^2 z \sqrt{g} \frac{\mathcal{Q}}{2} \mathcal{R} \mathrm{Re} Y$, which is necessary for conformal invariance of $\mathcal{N}=2$ Liouville theory. Note that the background charge $\mathcal{Q} = \frac{2}{k} $ is determined so that the $\mathcal{N}=2$ Liouville potential $\int d\theta^+ \mu Fe^{-Y}$ is the marginal deformation.

Let us discuss possible holomorphic (hence protected) F-term deformation of the $\mathcal{N}=2$ Liouville theory. An obvious deformation is the change of the $\mathcal{N}=2$ Liouville cosmological constant. This corresponds to the $(2,2)$ chiral deformation and it is dual to the marginal deformation \eqref{dudef}. The non-trivial F-term deformation is either given by the superpotential or non-trivial auxiliary field $E$ in the fermi multiplet. However, since the $\mathcal{N}=2$ Liouville theory only has one pair of chiral and fermi multiplet, the constraint $EJ = 0$ demands either $E=0$ or $J=0$, and as long as we keep $\mathcal{N}=2$ Liouville potential term, non-trivial $E$ deformation is impossible. Therefore, we can conclude that there is no holomorphic F-term deformation of the $\mathcal{N}=2$ Liouville field theory except for the change of the cosmological constant.

\section{Marginal $(0,2)$ deformation of two $\mathcal{N}=2$ Liouville theories and its dual}
In this section, we present an example of marginal $(0,2)$ deformation of non-compact Calabi-Yau space. For this purpose, we need two copies of $\mathcal{N}=2$ Liouville sector. The resultant theory has a non-trivial gauge bundle deformation from the $(2,2)$ locus as a conformal field theory.
\subsection{GLSM construction}
 We begin with the two copies of (generically different level $k$ and $\tilde{k}$) GLSM ($\Phi$, $P$, $\Gamma$, $\Gamma_P$, $\Upsilon$, $\Sigma$) and ($\tilde{\Phi}$, $\tilde{P}$, $\tilde{\Gamma}$, $\tilde{\Gamma}_P$, $\tilde{\Upsilon}$, $\tilde{\Sigma}$). The two systems are interacting through the choice of the auxiliary field
\begin{align}
E &=  \Sigma \Phi + \epsilon_1 \tilde{\Sigma} \Phi \cr
E_P &=  \Sigma + \epsilon_2\tilde{\Sigma} \cr
\tilde{E} &= \tilde{\Sigma}\tilde{\Phi} + \epsilon'_1\Sigma \tilde{\Phi} \cr
\tilde{E}_P &= \tilde{\Sigma} + \epsilon'_2\Sigma 
\end{align}
To obtain the exact dual superpotential without ambiguity, we restrict ourselves to the particular case with $\epsilon'_1 = \epsilon_2' = 0$ (see a discussion in the next subsection).

The component form of the Lagrangian is given by
\begin{align}
L &= -D^\mu \bar{\phi} D_{\mu}\phi + i \bar{\psi}_- D_+ \psi_- + i \bar\psi_+ D_- \psi_+ + D|\phi|^2  -|\sigma \phi + \epsilon_1 \tilde{\sigma} \phi|^2 \cr
&- \bar{\psi}_- \sigma \psi_+ - \bar{\sigma}_+ \bar{\sigma} \psi_- - \epsilon_1 \bar{\psi}_- \tilde{\sigma}\psi_+ - \epsilon_1 \bar{\psi}_+ \bar{\tilde{\sigma}} \psi_- \cr
&- i \bar{\phi} \lambda_- \psi_+ + i \bar{\phi}\lambda_+ \psi_- + i \bar{\psi}_+ \bar{\lambda}_- \phi - i \bar{\psi}_- \bar{\lambda}_+ \phi + \epsilon_1 i \bar{\phi} \tilde{\lambda}_+ \psi_- - \epsilon_1 i \bar{\psi}_- \bar{\tilde{\lambda}}_+ \phi \cr 
&+ \frac{k}{2}\left(-D^\mu\bar{p} D_\mu p + i\bar{\chi}_-\partial_+ \chi_- + i \bar{\chi}_+\partial_- \chi_+ + D(p+\bar{p}) - |\sigma+\epsilon_2\tilde{\sigma}|^2  \right. \cr
&+\left. i\chi_- \lambda_- + i \bar{\chi}_+\bar{\lambda}_-- i \chi_- \lambda_+ -i \bar{\chi}_- \bar{\lambda}_+ - i \epsilon_2 \chi_- \tilde{\lambda}_+ - i \epsilon_2 \bar{\chi}_-\bar{\tilde{\lambda}}_+ \right)\cr
 &- D^\mu \bar{\tilde{\phi}} D_{\mu} \tilde{\phi} + i \bar{\tilde{\psi}}_-D_+\tilde{\psi}_- +\tilde{D}|\tilde{\phi}|^2 +  i\bar{\tilde{\psi}}_+D_- \tilde{\psi}_+ \cr
&- |\tilde{\sigma}|^2|\tilde{\phi}|^2 - \bar{\tilde{\psi}}_- \tilde{\sigma} \tilde{\psi}_+ - \bar{\tilde{\psi}}_+ \bar{\tilde{\sigma}} \tilde{\psi}_- - i\bar{\tilde{\phi}} \tilde{\lambda}_- \tilde{\psi}_+ + i\bar{\tilde{\phi}} \tilde{\lambda}_+ \tilde{\psi}_- + i\bar{\tilde{\psi}}_+ \bar{\tilde{\lambda}}_- \tilde{\phi} - i \bar{\tilde{\psi}}_- \bar{\tilde{\lambda}}_+ \tilde{\phi} \cr
& + \frac{\tilde{k}}{2}\left(-(D_\mu \bar{\tilde{p}})(D^\mu \tilde{p}) + i\bar{\tilde{\chi}}_-\partial_+ \tilde{\chi}_- + i \bar{\tilde{\chi}}_+\partial_- \chi_+ + \tilde{D}(\tilde{p}+\bar{\tilde{p}})  \right. \cr
& \left. -|\tilde{\sigma}|^2 + i \tilde{\chi}_+ \tilde{\lambda}_- - i\tilde{\chi}_- \tilde{\lambda}+ + i \bar{\tilde{\chi}}_+\bar{\tilde{\lambda}}_- - i \bar{\tilde{\chi}}_- \bar{\tilde{\lambda}}_+ \right) \cr 
& + \frac{1}{2e^2}(-\partial^\mu \bar{\sigma}\partial_\mu \sigma + i\bar{\lambda}_- \partial_+ \lambda_- + i \bar{\lambda}_+ \partial_- \lambda_+ + v_{01}^2 + D^2) \cr
&+ \frac{1}{2\tilde{e}^2}(-\partial^\mu \bar{\tilde{\sigma}}\partial_\mu \tilde{\sigma} + i\bar{\tilde{\lambda}}_- \partial_+ \tilde{\lambda}_- + i \bar{\tilde{\lambda}}_+ \partial_- \tilde{\lambda}_+ + \tilde{v}_{01}^2 + \tilde{D}^2) \ .
\end{align}

To obtain the low energy effective action, we integrate out massive gauge multiplet $\Sigma$, $\tilde{\Sigma}$ first, which gives rise to the four-fermi interaction
{\footnotesize \begin{eqnarray}
\frac{\frac{k}{2}(\epsilon_1-\epsilon_2)(\bar{\psi}_-\psi_+ \bar{\tilde{\psi}}_+ \tilde{\psi}_- + \bar{\psi}_+\psi_- \bar{\tilde{\psi}}_- \tilde{\psi}_+) + \bar{\tilde{\psi}}_- \tilde{\psi}_+ \bar{\tilde{\psi}}_+ \tilde{\psi}_-(|\phi|^2 + \frac{k}{2}) + \bar{\psi}_- \psi_+ \bar{\psi}_+ \psi_-(|\tilde{\phi}|^2 + (\epsilon_1-\epsilon_2)^2\frac{k}{2} + \frac{\tilde{k}}{2})}{\frac{k}{2}(|\tilde{\phi}|^2 + \frac{\tilde{k}}{2}) + |\phi|^2(|\tilde{\phi}|^2 + (\epsilon_1 -\epsilon_2)^2 \frac{k}{2} + \frac{\tilde{k}}{2})} \ .  \label{ffm}
\end{eqnarray}}
We can see that at $(2,2)$ point, where $\epsilon_1=\epsilon_2$, the gauge bundle is just given by the sum of the tangent bundle: $V = TM_1 \oplus TM_2$, while there is a non-trivial mixing for general deformation parameters as can be seen from non-zero $\bar{\psi}_-\psi_+ \bar{\tilde{\psi}}_+ \tilde{\psi}_- $ term. Integrating out gauginos $\lambda_{\pm}$ and $\tilde{\lambda}_{\pm}$ gives the relation
\begin{align}
\frac{k}{2}\chi_+ &= -\bar{\phi} \psi_+ \cr
\frac{k}{2}\chi_- &= -\bar{\phi} \psi_- \cr
\frac{\tilde{k}}{2} \tilde{\chi}_+ &= -\bar{\tilde{\phi}} \tilde{\psi}_+ \cr
\frac{\tilde{k}}{2} \tilde{\chi}_- &= -\bar{\tilde{\phi}} \tilde{\psi}_- - (\epsilon_1-\epsilon_2) \bar{\phi} \psi_-  \ . \label{rrs}
\end{align}
Note that only the left-moving fermion is modified due to the deformation.

Furthermore, we can integrate out $\Upsilon$, $\tilde{\Upsilon}$ multiplets by fixing the gauge $\mathrm{Im} p = \mathrm{Im} \tilde{p} = 0 $, and solving the $D$-term condition as $|\phi|^2 = -k \mathrm{Re} p $, $|\tilde{\phi}|^2 = - \tilde{k} \mathrm{Re} \tilde{p}$. The bosonic part of the action is not deformed and is given by the sigma model on two distinct manifolds:
\begin{eqnarray}
ds^2 = (1+ \frac{r^2}{k}) dr^2 + \frac{d\theta^2}{(1+\frac{r^2}{k})} + (1+ \frac{\tilde{r}^2}{\tilde{k}}) d\tilde{r}^2 + \frac{d\tilde{\theta}^2}{(1+ \frac{\tilde{r}^2}{\tilde{k}})} \ ,
\end{eqnarray}
where $\phi = \frac{r}{\sqrt{2}} e^{i\theta}$ and $\tilde{\phi} = \frac{\tilde{r}}{\sqrt{2}} e^{i\tilde{\theta}}$.

Similarly, the fermionic part of the action can be obtained as
\begin{align}
L_F  &= i\bar{\psi}_- \partial_+ \psi_- + i \bar{\psi}_+ \partial_- \psi_+ + \frac{ik}{2}\bar{\chi}_- \partial_+ \chi_- + \frac{ik}{2}\bar{\chi}_+ \partial_- \chi_+  \cr
& + \frac{\bar{\psi}_- \psi_-}{\frac{k}{2}+|\phi|^2} \left(\frac{i}{2}\phi \partial_+ \bar{\phi} - \frac{i}{2}\bar{\phi}\partial_+\phi \right) + \frac{\bar{\psi}_+\psi_+}{\frac{k}{2}+|\phi|^2} \left(\frac{i}{2}\phi\partial_-\bar{\phi}-\frac{i}{2}\bar{\phi}\partial_-\phi\right)   \cr
& - \frac{1}{\frac{k}{2} + |\phi|^2} (\bar{\psi}_-\psi_-\bar{\psi}_+\psi_+) \cr
&+ i\bar{\tilde{\psi}}_- \partial_+ \tilde{\psi}_- + i \bar{\tilde{\psi}}_+ \partial_- \tilde{\psi}_+ + \frac{i\tilde{k}}{2}\bar{\tilde{\chi}}_- \partial_+ \tilde{\chi}_- + \frac{i\tilde{k}}{2}\bar{\tilde{\chi}}_+ \partial_- \tilde{\chi}_+  \cr
& + \frac{\bar{\tilde{\psi}}_- \tilde{\psi}_-}{\frac{\tilde{k}}{2}+|\tilde{\phi}|^2} \left(\frac{i}{2}\tilde{\phi} \partial_+ \bar{\tilde{\phi}} - \frac{i}{2}\bar{\tilde{\phi}}\partial_+\tilde{\phi} \right) + \frac{\bar{\tilde{\psi}}_+\tilde{\psi}_+}{\frac{\tilde{k}}{2}+|\tilde{\phi}|^2} \left(\frac{i}{2}\tilde{\phi}\partial_-\bar{\tilde{\phi}}-\frac{i}{2}\bar{\tilde{\phi}}\partial_-\tilde{\phi}\right)   \cr
& - \frac{1}{\frac{\tilde{k}}{2} + |\tilde{\phi}|^2} (\bar{\tilde{\psi}}_-\tilde{\psi}_-\bar{\tilde{\psi}}_+\tilde{\psi}_+)
\end{align}
together with the additional four-fermi term \eqref{ffm}. We can see again that the gauge bundle is only modified through the substitution of \eqref{rrs}. In particular, connection for the right-mover is not modified and is compatible with the Riemann metric as it should be.

The first order (off-diagonal) deformation of the gauge connection with respect to $\epsilon_1-\epsilon_2$ is 
\begin{eqnarray}
\delta A = (\epsilon_1-\epsilon_2) \frac{i}{k}  \frac{({\tilde{\phi}} d \bar{\phi} - \phi d \bar{\tilde{\phi}})}{\sqrt{1+\frac{2}{k}|\phi|^2}\sqrt{1+\frac{2}{\tilde{k}}|\tilde{\phi}|^2}}  \ .
\end{eqnarray}
Here, the connection is a deformation of the vector bundle for the second manifold from the standard embedding $V=TM_2$ (i.e. the introduction of non-trivial $A_{\psi,\tilde{\psi}}$).

Before turning on the $(0,2)$ deformation, the low energy effective field theory of the GLSM is given by the direct sum of the two $\mathcal{N}=2$ $SL(2,\mathbf{R})/U(1)$ coset models (with level $k$ and level $\tilde{k})$. The non-trivial $(0,2)$ deformation mixes the two coset models, and in contrast to the example discussed in the last section, this induces a non-trivial deformation even at the conformal fixed point.

In the low energy coset description, the deformation corresponds to
\begin{eqnarray}
[{J}_{-1}^-|j=\frac{k}{2}+1\rangle^+ \otimes |{\psi}_{-\frac{1}{2}}|0 \rangle]^{w=-1}  \otimes [\tilde{J}_0^+|j=\frac{\tilde{k}}{2}\rangle^+ \otimes |\bar{\tilde{\psi}}_{-\frac{1}{2}}|0 \rangle]^{w=-1} \ , \label{defl}
\end{eqnarray}
which is $(1,1)$ deformation that preserves the $(0,2)$ supersymmetry.\footnote{The $\tilde{}$ notation here has a double meaning: the one is the right-mover and the other is the second $SL(2,\mathbf{R})/U(1)$ coset with level $\tilde{k}$. } To see this, we simply note that the right-mover is the same as the $(2,2)$ deformation, but the left-mover breaks the half amount of supersymmetry. In the next section, we also find the corresponding deformation in the dual $\mathcal{N}=2$ Liouville theory from the non-perturbative instanton contributions to the dual superpotential.

\subsection{dual theory}
The perturbative duality gives the following kinetic terms
\begin{align}
L &= \int d^2\theta\left[ \frac{i}{8}\frac{Y-\bar{Y}}{Y+\bar{Y}}\partial_-(Y + \bar{Y}) + \frac{i}{k} \bar{Y}_P \partial_- Y_P - \frac{\bar{F}F}{Y+\bar{Y}} - \frac{1}{k}\bar{F}_P F_P \right] \cr
 & + \int d^2\theta\left[ \frac{i}{8}\frac{\tilde{Y}-\bar{\tilde{Y}}}{\tilde{Y}+\bar{\tilde{Y}}}\partial_-(\tilde{Y} + \bar{\tilde{Y}}) + \frac{i}{\tilde{k}} \bar{\tilde{Y}}_P \partial_- \tilde{Y}_P - \frac{\bar{\tilde{F}}\tilde{F}}{\tilde{Y}+\bar{\tilde{Y}}} - \frac{1}{\tilde{k}}\bar{\tilde{F}}_P \tilde{F}_P \right]\
\end{align}
together with the perturbative superpotential
\begin{eqnarray}
W = -\frac{i\Upsilon}{4}(Y+2Y_P) -\frac{i\tilde{\Upsilon}}{4}(\tilde{Y}+2\tilde{Y_P}) +\frac{\Sigma (F+F_P)}{\sqrt{2}} + \frac{\tilde{\Sigma}(\tilde{F} + \tilde{F}_P + \epsilon_1 F + \epsilon_2 F_P)}{\sqrt{2}} \ . 
\end{eqnarray}

The structure of the non-perturbative superpotential with general deformation $(\epsilon_1, \epsilon_2, \tilde{\epsilon}_1, \tilde{\epsilon}_2)$ would be
\begin{eqnarray}
W_{\mathrm{non-pert}} = \beta_1 F e^{-Y} + \beta_2 \tilde{F} e^{-Y} + \tilde{\beta}_1 F e^{-\tilde{Y}} + \tilde{\beta}_2 e^{-\tilde{Y}} \ ,
\end{eqnarray}
where we have assumed that there is no contribution from the axion multiplets. At $(2,2)$ point, where $\epsilon_i = \tilde{\epsilon}_i$, we have $\beta_1 = \tilde{\beta}_2 = \mu$, $\beta_2 = \tilde{\beta}_1 = 0$. To study the deformation further, we introduce the following (spurious) symmetry: $Q(\Sigma) = k$, $Q(\tilde{\Sigma}) = \tilde{k}$, under which the deformation parameters are charged with $Q(\epsilon_1) = Q(\epsilon_2) = k -\tilde{k}$, and $Q(\epsilon'_1) = Q(\epsilon'_2) = \tilde{k}-k$. 
Under the symmetry, $\epsilon_1 \epsilon_1', \epsilon_1\epsilon_2', \epsilon_2 \epsilon_1', \epsilon_2 \epsilon_2'$ are not charged, so arbitrary powers of these combination could appear in the dual action. To obtain unambiguous dual action, we have assumed $\epsilon_1' = \epsilon_2' = 0$.

Now, in the dual variable, the spurious symmetry gives $Q(F) = -k$ and $\tilde{F} = - \tilde{k}$. Furthermore, since the $(2,2)$ dual action should be invariant, we have $Q(e^{-Y}) = k$ and $Q(e^{-\tilde{Y}}) = \tilde{k}$. From the invariance of the dual action, we can read the charge of dual parameter $Q(\beta_1) = Q(\tilde{\beta}_2)= 0$, $Q(\beta_2) = \tilde{k}-k$, and $Q(\tilde{\beta}_1) = k -\tilde{k}$. The continuity at $\epsilon_i=0$ uniquely determines the $\epsilon_i$ dependence on $\beta$: $\beta_1 = \mu$, $\beta_2 = 0$, $\tilde{\beta}_1 = a(\epsilon_1 - \epsilon_2)$, $\tilde{\beta}_2 = \tilde{\mu}$. Therefore, the dual superpotential is finally given by
\begin{eqnarray}
W_{\mathrm{non-pert}} = \mu F e^{-Y} + a (\epsilon_1 - \epsilon_2) F e^{-\tilde{Y}} + \tilde{\mu} \tilde{F} e^{-\tilde{Y}} \ .
\end{eqnarray}
Note that proportionality with $\epsilon_1-\epsilon_2$ is consistent with vanishing coefficient at $(2,2)$ point. Note that even if this term had not arise from the instanton effect, it would appear effectively after integrating out massive fields as we will see.

To study the low-energy physics, we integrate out the massive gauge multiplets, giving the constraint
\begin{align}
Y_P &= -\frac{1}{2}Y \ \ \ \ \tilde{Y}_P = -\frac{1}{2}\tilde{Y} \cr
F_P &= -F \ \ \ \ \ \ \tilde{F}_P = -\tilde{F} -(\epsilon_1-\epsilon_2) F \ . 
\end{align}
In order to obtain a canonical kinetic term for large $\mathrm{Re} \tilde{Y}$, we redefine $\tilde{F} + (\epsilon_1 - \epsilon_2) F \to \tilde{F}$. Then, we have the effective kinetic term (for large $\mathrm{Re} \tilde{Y}$)
\begin{eqnarray}
L = \int d^2\theta \frac{i}{4k} \bar{Y}\partial_- Y -\frac{1}{k} \bar{F}F + \frac{i}{4\tilde{k}} \bar{\tilde{Y}}\partial_- \tilde{Y} -\frac{1}{\tilde{k}} \bar{\tilde{F}}\tilde{F} \ , 
\end{eqnarray}
with the effective superpotential
\begin{eqnarray}
W_{\mathrm{non-pert}} = \mu F e^{-Y} + \tilde{a} (\epsilon_1 - \epsilon_2) F e^{-\tilde{Y}} + \tilde{\mu} \tilde{F} e^{-\tilde{Y}} \ ,
\end{eqnarray}
where $\tilde{a}$ is shifted from $a$ due to the redefinition $\tilde{F} + (\epsilon_1-\epsilon_2)F \to \tilde{F}$ just mentioned above. This is the final form of our proposed dual action describing the $(0,2)$ deformation of the $SL(2,\mathrm{R})/U(1)$ coset models. The new $(0,2)$ Liouville interaction $W= \tilde{a} (\epsilon_1 - \epsilon_2) F e^{-\tilde{Y}}$ just corresponds to \eqref{defl}.

To make the story complete, let us discuss possible $F$-term (holomorphic) deformations of two $\mathcal{N}=2$ Liouville theories. $(2,2)$ deformation should be given by the $(2,2)$ superpotential $ W_{(2,2)} =e^{ n b_1 S_1 + m b_2 S_2}$. The compactification of the imaginary part of the Liouville field\footnote{From the purely $\mathcal{N}=2$ Liouville theory viewpoint, this is not necessary at all. However, the duality to $SL(2,\mathbf{R})/U(1)$ coset model demands the quantization, and we only focus on these cases.} $S_1$ and $S_2$ suggest that $n$ and $m$ should be integers. On the other hand, the marginality condition gives
\begin{eqnarray}
n + m = 1 \ ,
\end{eqnarray}
so there is no non-trivial solution except for the original $\mathcal{N}=2$ Liouville potential $(n,m) = (1,0)$ or $(0,1)$ due to the unitarity constraint.

$(0,2)$ deformation comes from changing the superpotential by $J^a$ or changing the auxiliary field for the fermi multiplet by $E^a$. A Similar argument above shows that the possible $(0,2)$ deformation from the $(0,2)$ superpotential (we decompose the $(2,2)$ chiral mulitiplet $S$ into a $(0,2)$ chiral multiplet $\Phi$ and a Fermi multiplet $F$) is given by
\begin{eqnarray}
 F_2 e^{b_1\Phi_1}  \ , \ \ \ F_1 e^{b_2 \Phi_2} \ , \label{sud}
\end{eqnarray}
which is just the dual for the deformation studied in section 4.\footnote{A similar supersymmetry breaking fermionic deformation was studied in \cite{Nakayama:2006gt} in the cosmological context.} Any other dual operators violate unitarity.

If we turned off the $(0,2)$ deformation \eqref{sud} from the superpotential, it would seem possible to introduce non-trivial $E_a$ deformations. $E_a$ should satisfy the supersymmetry condition
\begin{eqnarray}
E_1 \mu_1 e^{b_1\Phi_1} + E_2 \mu_2 e^{b_2\Phi_2} = 0 \ .
\end{eqnarray}
Furthermore, the marginality condition and the quantization of the Liouville exponent uniquely fixes $E_a$ as
\begin{eqnarray}
E_1 = \epsilon \mu_2 e^{b_2\Phi_2} \ ,  \ \ E_2 = -\epsilon \mu_1 e^{b_1\Phi_1} \ .
\end{eqnarray}
However, it is not difficult to see that all the induced interaction such as
\begin{eqnarray}
\delta L = - \epsilon \mu_2 b_2 \bar{\chi}_{-,1} \psi_{+,2} e^{b_2 \Phi_2} + \epsilon \mu_1 b_1 \bar{\chi}_{-,2} \psi_{+,1} e^{b_1\Phi_1} + h.c. \ 
\end{eqnarray}
 is trivially removed by the field redefinition of the right hand fermions
\begin{align}
\chi_{-,1} &\to \chi_{-,1} + \epsilon \bar{\chi}_{-,2} \cr
\chi_{-,2} &\to \chi_{-,2} - \epsilon \bar{\chi}_{-,1} \ .
\end{align}
Thus, we conclude that there is no non-trivial $F$-term deformation of the two $\mathcal{N}=2$ Liouville theories (for general $k$ and $\tilde{k}$) except for the ones discussed in this section.

\section{Geometric interpretation}
In this section, we give geometric interpretations of the duality so far obtained in previous sections.
\subsection{$\mathcal{N}=2$ Liouville theory and non-compact Calabi-Yau}
The $\mathcal{N}=2$ Liouville theory has geometrical interpretations as non-compact Gepner model constructions of the non-compact Calabi-Yau spaces (see e.g.  \cite{Lerche:2000uy}\cite{Eguchi:2004ik}). A classical example of the non-compact Gepner model constructions would be Ghoshal-Vafa duality between $\mathcal{N}=2$ Liouville theory ($\mathcal{N}=2$ $SL(2,\mathbf{R})/U(1)$ coset model) at $k=1$ and the deformed conifold background \cite{Ghoshal:1995wm}. We have seen in section 4 that the non-trivial vector bundle deformation of the heterotic string on the deformed conifold cannot be studied from the simple $(0,2)$ deformation of the GLSM.

In section 5, in contrast, we studied non-trivial $(0,2)$ deformation of two $\mathcal{N}=2$ Liouville theories, which presumably corresponds to deformations of the gauge bundle moduli on the dual non-compact Calabi-Yau spaces. We can embed these theories in string theory as a non-compact Gepner construction. We recall that the criticality condition of the string theory demands
\begin{eqnarray}
\left(1+\frac{2}{k}\right) + \left(1 + \frac{2}{\tilde{k}}\right) = n \ ,
\end{eqnarray}
for Calabi-Yau $n$-fold. The simplest example would be $k=\tilde{k}=4$ for Calabi-Yau 3-fold which describes $ALE(A_1)$ fibration over $CP_1$. 

To discuss the corresponding geometry further, we recall the Calabi-Yau/Landau-Ginzburg correspondence: the Calabi-Yau $n$-fold defined by 
\begin{eqnarray}
X_1^{r_1} + \cdots + X_{n+2}^{r_{n+2}} = 0 , \ \ \mathrm{in} \ WCP_{n+1}\left(\frac{1}{r_1},\cdots, \frac{1}{r_{n+2}}\right)
\end{eqnarray}
with $\sum_{i=1}^{n+2}\frac{1}{r_i} = 1$ is equivalent to the Landau-Ginzburg orbifold with the $(2,2)$ superpotential $W_{(2,2)}(X_i) = X_1^{r_1} + \cdots + X_{n+2}^{r_{n+2}}$. As we will see, our case with non-compact Calabi-Yau space requires that some of the power $r_i$ be negative, and the Landau-Ginzburg description is rather formal at this stage \cite{Giveon:1999zm}\cite{Lerche:2000uy}\cite{Hori:2002cd}\cite{Eguchi:2004ik}. 

The non-compact version of the Calabi-Yau/Landau-Ginzburg correspondence goes in the following way.  Let us consider the Landau-Ginzburg model with the $(2,2)$ superpotential
\begin{eqnarray}
W = X_1^2 + X_2^2 + X_3^2 + Y_1^{-k} + Y_2^{-\tilde{k}} \ ,
\end{eqnarray}
where $\frac{1}{k}+\frac{1}{\tilde{k}} = \frac{1}{2}$, corresponding to a non-compact Calabi-Yau 3-fold
\begin{eqnarray}
X_1^2 + X_2^2 + X_3^2 + Y_{1}^{-k} + Y_2^{-\tilde{k}} = 0 \ , \ \ \mathrm{in} \ WCP_4\left(k\tilde{k}, k\tilde{k}, k\tilde{k}, -2\tilde{k}, -2k \right) \ .
\end{eqnarray}
To make sense of the negative power in the superpotential and gain more geometrical intuition of the target space, we introduce the Liouville coordinate \cite{Giveon:1999zm}
\begin{eqnarray}
Y_1^{-k} = e^{-\sqrt{\frac{k}{2}}\Phi_1} \ \ Y_2^{-\tilde{k}} = e^{-\sqrt{\frac{\tilde{k}}{2}}\Phi_2} \ .
\end{eqnarray}
The Jacobian of the path integral associated with this change of variables induces a linear dilaton factor (see e.g. \cite{Nakayama:2004at})
\begin{eqnarray}
\Phi = -\sqrt{\frac{1}{2k}}\mathrm{Re} \Phi_1 - \sqrt{\frac{1}{2\tilde{k}}}\mathrm{Re} \Phi_2 \ .
\end{eqnarray}
Now the theory is well behaved as a sum of two $\mathcal{N}=2$ Liouville theories. 

Similarly one can rewrite the superpotential as 
\begin{eqnarray}
W = e^{-nZ}\left(e^{Y/k_1} + e^{Y/k_2} + X_1^2 + X_2^2 + X_3^2 \right) \ ,
\end{eqnarray}
and integrate out $Z$ field, resulting in the geometry
\begin{eqnarray}
e^{Y/k_1} + e^{Y/k_2} + X_1^2 + X_2^2 + X_3^2 = 0
\end{eqnarray}
describing the $ALE(A_1)$ fibration over $WCP_1(k_1,k_2)$.

As a particular example, we take $n=4$, $k=\tilde{k} = 2$, which has a direct geometrical construction studied  in the literature. The model is given by two copies of $ALE(A_1)$ space, or $O(-2)\oplus O(-2)$ bundle over $CP_1 \times CP_1$ with  further vector bundle deformations. Actually, the vector bundle deformation of this model can be analysed by using a different GLSM from us (without any axionic matter) as has been done in \cite{Adams:2003zy}. In their model, they introduced $U(1)_1 \times U(1)_2$ with two charge one chiral multiplets $\Phi_1,\Phi_2$ ( $\tilde{\Phi}_2, \tilde{\Phi}_2$ for $U(1)_2$) and charge $-2$ chiral multiplet $P$ (and $\tilde{P}_2$). After integrating out massive multiplets (dual of $\Phi_i$ and $\tilde{\Phi}_i$), it is not difficult to see that our effective superpotential after duality completely agrees with the one studied in \cite{Adams:2003zy}:\footnote{One should be careful, however, because the authors of \cite{Adams:2003zy} did a coordinate transformation to make the Liouville directions compact and treated them as if it were a conventional Landau-Ginzburg model. The non-compactness of the target space is not manifest in their approach and we believe that a physically suitable coordinate involves Liouville directions as we have done. In addition, some of the instanton parameters were not fixed in \cite{Adams:2003zy}, and the consistency to our approach should give a constraint on their exact parameter map.} the vector bundle deformation is described by the two Liouville field theory with the $(0,2)$ superpotential $ W_{(0,2}) = F e^{-Y} + \tilde{F} e^{-\tilde{Y}} + \epsilon F e^{-\tilde{Y}}$. 

An important consequence of this construction is that one could (in principle) read the geometric data of the vector bundle deformation from the parent GLSM corresponding to our Liouville deformation. Mathematically, the vector bundle deformation in conventional heterotic compactifications is described by $H_1(M,End(V))$ and might be computed explicitly from the GLSM. One problem, however, is that the classical GLSM does not give a Calabi-Yau metric nor the vector bundle deformation consistent with the heterotic equations of motion (hence it is not conformal at one-loop). The study of the renormalization group equation would yield a conformal fixed point, but the actual computation is cumbersome and furthermore we may still have to deal with non-perturbative effects. The good point of our dual formulation based on the $\mathcal{N}=2$ Liouville theory (or $SL(2,\mathbf{R})/U(1)$ coset model) is that the conformal property is manifest and some important quantities are not renormalized due to the holomorphic nature of the superpotential.

\section{Discussion}
In this paper, we have studied the mirror duality of the $(0,2)$ non-compact Calabi-Yau space with non-trivial gauge bundle deformations. Our approach has been a composition of the effective field theory analysis from the non-linear sigma model and the world-sheet exact analysis based on the Liouville theory and coset model. The former has given us the intuitive geometric understanding of the duality, while the latter knows exactly the (ir-)relevance of the geometric deformation at the quantum level.

The FZZ duality itself can be seen as a duality between the tachyon condensation (sine-Liouville phase) and the geometric resolution of singularity (2D black hole phase). The world-sheet non-perturbative corrections show different aspects in each phase, but the physics is the same if we quantize the system exactly. The world-sheet exact treatment (solvability of the Liouville theory) here plays a significant role because the full quantum corrections are under control. In this paper, we have only discussed the small perturbation around the (2,2) background from the exact CFT viewpoint, but it would be very interesting see if the solvability continues to hold away from the (2,2) point. Various techniques used in the Liouville theory (see \cite{Nakayama:2004vk} for a review) may remain useful here.

\section*{Acknowledgements}
The research of Y.~N. is supported in part by NSF grant PHY-0555662 and the UC Berkeley Center for Theoretical Physics. A part of this work was completed during Summer Institute 2008 at Fuji Yoshida, and the author thanks all the participants and organizers. He especially acknowledges the fruitful discussions on the subject there with T.~Eguchi and S.~Yamaguchi.

\appendix

\section{$\mathcal{N}=2$ $SL(2,\mathbf{R})/U(1)$ coset model}
In this appendix, we review some basic aspects of $\mathcal{N}=2$ $SL(2,\mathbf{R})/U(1)$ Kazama-Suzuki coset model \cite{Kazama:1988qp}\cite{Kazama:1988uz}. 
We begin with the bosonic $SL(2,\mathbf{R})$ WZNW model.
It is generated by the world-sheet current\footnote{When we talk about conformal field theories, we use $\tilde{}$ to denote the right-mover compared with $\tilde{}$-less expression for the left-mover. We hope this will not be confusing.}
\begin{eqnarray}
J_{L}^a = \sum_{n=-\infty}^{\infty} J_n^a e^{iny^-} \ , \ \ J_R^a = \sum_{n=-\infty}^{\infty} \tilde{J}_n^a e^{in y^+} \ .
\end{eqnarray}
The commutation relations are
\begin{align}
[J_n^{3},J_m^{3}] &= -\frac{k_B}{2} n \delta_{n+m,0} \cr
[J_n^{3},J_m^{\pm}] &= \pm J_{n+m}^{\pm} \cr
[J_n^{+},J_m^{-}] &= -2 J_{n+m}^{3} + k_B n \delta_{n+m,0} \ ,
\end{align}
where $k_B$ is the (bosonic) level of the current algebra.

The supersymmetric $SL(2,\mathbf{R})$ WZNW model is described by bosonic $SL(2,\mathbf{R})$ WZNW model with $k_B = k+2$ with three free fermions.\footnote{In the Kazama-Suzuki coset, only two fermions (= Dirac fermion) out of three, $\psi = \psi^1 + i\psi^{2} $ and $\bar{\psi} = \psi^{1}-i\psi^{2}$ are important. The other $\psi^{3}$ would be eliminated through the coset construction.} The fermion is charged under the total $SL(2,\mathbf{R})$ algebra with the commutation relation
\begin{align}
[J_0^{3(t)},\psi_r] = -\psi_r \ , \ \ [J_0^{3(t)}, \bar{\psi}_r] = \bar{\psi}_r \cr
[\tilde{J}_-^{3(t)}, \tilde{\psi}_r] = \tilde{\psi}_r \ , \ \ [\tilde{J}^{3(t)}_0, \bar{\tilde{\psi}}_r] = -\bar{\tilde{\psi}}_r \ . 
\end{align}
In other worlds, the total $SL(2,\mathbf{R})$ current is given by the sum of the bosonic part $J^{a(b)}$ and the fermionic part $\psi\bar{\psi}$.

The Hilbert space of the supersymmetric $SL(2,\mathbf{R})$ WZNW model is given by the direct product of bosonic $SL(2,\mathbf{R})_{k+2}$ WZNW model and the Fock space of the Dirac fermion. A part of the former is obtained from the following representations of $SL(2,\mathbf{R})$ as the Kac-Moody primaries \cite{Maldacena:2000hw}\cite{Hori:2002cd}
\begin{enumerate}
	\item $\mathcal{D}_j^+$: principal discrete representation with lowest weight (i.e. $j^3 \equiv m = j , j+1 ,j+2 \cdots$) of spin $j$, where $\frac{1}{2} < j < \frac{k+1}{2} $ \ .
\item $\mathcal{D}_j^-$: principal discrete representation with highest weight (i.e. $j^3 \equiv m = -j , -j-1 ,-j-2 \cdots$) of spin $j$, where $\frac{1}{2} < j < \frac{k+1}{2} $ \ .
\item $\mathcal{C}^\alpha_j$: principal continuous representations with $j=\frac{1}{2}+ ip$ , $p\in \mathbf{R_{\ge 0}}$ and $0 \le \alpha < 1$ ($j^3\equiv m = \alpha, \alpha \pm 1, \alpha \pm2 \cdots$).
\end{enumerate}
We denote the condition $\frac{1}{2}<j<\frac{k+1}{2}$ for discrete representations as the unitarity condition \cite{Maldacena:2000hw}.
The corresponding Kac-Moody primaries are denoted by $\hat{\mathcal{D}}_{j}^{\pm}$ and $\hat{\mathcal{C}}^{\alpha}_j$. 
We recall $J_n^{a}$ annihilate Kac-Moody primaries for all $n>0$. They have the conformal weights
\begin{eqnarray}
L_0 = \bar{L}_0 = -\frac{j(j-1)}{k_B-2} \ .
\end{eqnarray}

In addition, we include spectral flowed representations of these basic representations \cite{Maldacena:2000hw}. The spectral flow automorphism of the current algebra is obtained by $J_n^{a} \to \hat{J}^{a}_n$ with
\begin{eqnarray}
\hat{J}_n^3 = J_n^3 -\frac{k_B}{2}w\delta_{n,0}\ ,  \ \ \hat{J}^+_n = J^{+}_{n+w} \ ,  \ \ \hat{J}^{-}_n = J^{-}_{n-w} \ ,
\end{eqnarray}
where $w \in \mathbf{Z}$ is the amount of spectral flow. In particular, the quantum number of $L_0$ and $J_0^3$ changes as $(h,m) \to (h+wm-\frac{k_Bw^2}{4}, m-{k_Bw}/2)$. In the supersymmetric theory, the spectral flow also acts on the Dirac fermion. It sends the fermion Fock space to itself. For example
\begin{eqnarray}
|0 \rangle \to \bar{\psi}_{-w+\frac{1}{2}} \cdots \bar{\psi}_{-\frac{1}{2}} \tilde{\psi}_{-w+\frac{1}{2}} \cdots \tilde{\psi}_{-\frac{1}{2}}|0 \rangle \ ,
\end{eqnarray}
for $w \ge 1$ under the spectral flow $-w$. Total quantum number, therefore is transformed as
\begin{align}
J_0^{3(t)} &= m -\frac{kw}{2} \ , \ \  \tilde{J}^3_0 = \tilde{m} -\frac{kw}{2} \cr
L_0 &= -\frac{j(j-1)}{k} + w m - \frac{k}{4}w^2 \ , \ \ \tilde{L}_0 = -\frac{j(j-1)}{k} + w \tilde{m} - \frac{k}{4} w^2 \ .
\end{align}
We note that the amount of the spectral flow should be the same  both for the left-mover and the right-mover.

In the coset theory, states are restricted by the gauging condition $J^{3(t)}_0 + \tilde{J}^{3(t)}_0 =  0$ and $J_n^3 = \tilde{J}_n^3 = 0$ for $n\ge 0$.  We define the momentum quantum number\footnote{The momentum $n$ is quantized: $n \in \mathbf{Z}$.} by $ n \equiv J^{3(b)}_0 - \tilde{J}^{3(b)}_0$, where $J^{3(b)}_0$ is the bosonic part of the $SL(2,\mathbf{R})$ generator $J_0^{3}$. Under the coset construction of the Virasoro generator: $T^{SL(2,\mathbf{R})/U(1) } = T^{SL(2,\mathbf{R})} - T^{U(1)} $, we obtain, in particular,
\begin{align}
L_0 &= -\frac{j(j-1)}{k} + \frac{(m+s)^2}{k} + \frac{s^2}{2} \cr
\bar{L_0} &= -\frac{j(j-1)}{k} + \frac{(\bar{m}+\bar{s})^2}{k} + \frac{\bar{s}^2}{2} \ ,
\end{align}
where $m = \frac{n - kw}{2}$, and $\bar{m} = -\frac{n + kw}{2}$ for $\mathcal{N}=2$ primary operators (i.e. annihilated by $G^+_{r\ge \frac{1}{2}-s}$ and $G^{-}_{r \ge \frac{1}{2} +s}$). Here $s$ denotes the fermionic spin and $s=0$ corresponds to the NS vacuum. Other coset states are created over these primary operators by acting $J^{\pm}$, $\psi$ and $\bar{\psi}$ oscillators.\footnote{As usual, $J^{3}$, $\psi^3$ oscillators and ghost oscillators practically do not contribute because they are projected out by the BRST procedure of the gauging.}
The spectrum of the $\mathcal{N}=2$ $SL(2,\mathbf{R})/U(1)$ coset model can be also read from the partition function. See e.g. \cite{Eguchi:2004yi}\cite{Israel:2004ir} for details.

The coset theory possesses an enhanced $(2,2)$ supersymmetry algebra.
It is  generated by 
\begin{align}
G_r &= \sqrt{\frac{2}{k}}\sum_{n} \psi_{r+n} J_{-n}^+ \cr
\bar{G}_r &= \sqrt{\frac{2}{k}} \sum_{n} \bar{\psi}_{r+n} J_{-n}^- \cr
J_n &= \frac{1}{k+2} J_n^{3(b)} + \frac{1}{2} J_n^{3(f)} \ ,
\end{align}
for left-mover and 
\begin{align}
\tilde{G}_r &= \sqrt{\frac{2}{k}}\sum_{n}\tilde{\psi}_{r+n} \tilde{J}_{-n}^- \cr
\tilde{\bar{G}}_r &= \sqrt{\frac{2}{k}} \sum_{n} \bar{\tilde{\psi}}_{r+n} \tilde{J}_{-n}^+ \cr
\tilde{J}_n &= \frac{1}{k+2} \tilde{J}_n^{3(b)} + \frac{1}{2} \tilde{J}_n^{3(f)} \ ,
\end{align}
for right-mover. The commutation relation is 
\begin{align}
[L_m, G_r] &= \left(\frac{m}{2}- r\right) G_{m+r} \cr
[L_m, J_n ] &= - nJ_{m+n} \cr
\{G_r, \bar{G}_s\} &= 2L_{r+s} + (r-s)J_{r+s} + \frac{c}{3}\left(r^2-\frac{1}{4}\right)\delta{r,-s} \cr
\{G_r, G_s\} &= 0 \cr
[J_n, G_r] &= G_{r+n} \ \ \ \ [J_n,\bar{G}_r] = -\bar{G}_{r+n} \cr
[J_m,J_n] &= \frac{c}{3}m \delta_{m,-n} \ ,
\end{align}
where $c= 3 + \frac{6}{k}$.

We turn to the statement made in the main text. The claim is there is no $(0,2)$ supersymmetric marginal deformation of the $\mathcal{N}=2$ $SL(2,\mathbf{R})/U(1)$ coset model. For this purpose, we have to look for $(1,1)$ primary operators annihilated by half of the supercharge up to total derivatives. We begin with discrete representations and their spectral flow $\hat{\mathcal{D}}^{+,w}_j \otimes \hat{\mathcal{D}}^{+,w}_j$. The $(1,1)$ condition becomes
\begin{align}
L_0 &= - \frac{j(j-1)}{k} + \frac{(\frac{n-kw}{2} + s)^2}{k} + \frac{s^2}{2} + N = 1 \cr
\bar{L}_0 &= - \frac{j(j-1)}{k} + \frac{(-\frac{n+kw}{2} + \bar{s})^2}{k} + \frac{\bar{s}^2}{2} + \bar{N} = 1 \ , 
\end{align}
where $N , \bar{N} \in \mathbf{Z}_{\ge 0} $ are contribution from oscillators. In addition, we have a highest weight condition
\begin{align}
m &= \frac{n-kw}{2} = j + q \cr
\bar{m} &= -\frac{n+kw}{2} = j + \bar{q} 
\end{align}
with $q , \bar{q} \in \mathbf{Z}$, where $q$ counts number of $J^+_{-n \le 0}$ minus number of $J^{-}_{-n <0}$ and similarly for $\bar{q}$. As in the main text, we require that this mass-shell condition apply for all $k$, which result in four equations
\begin{align}
-\frac{n}{2}(2q + 1 ) - q(q+1) + ns + s^2 &= 0 \cr
\frac{w}{2}(2q + 1) - sw + \frac{s^2}{2} + N &= 1 \cr
\frac{n}{2}(2q+ 1) - q(q+1) - n \bar{s} + \bar{s}^2 &= 0 \cr
\frac{w}{2}(2q+1) - \bar{s} w + \frac{\bar{s}^2}{2} + \bar{N} &= 1 \ .
\end{align}

We restrict ourselves to the case with NS states $ s,\bar{s} \in \mathbf{Z}$. In this case, furthermore, we set $s = \bar{s} = 0$ and create fermionic states with explicit oscillators (counted by $N, \bar{N} \in \frac{1}{2} \mathbf{Z}_{\ge 0} $). It is easy to see that the condition reduces to the case $n=0$ and $q = \bar{q} = 0, -1$ with $ \pm \frac{w}{2} + N =1$. The unitarity condition further sets $w = 0, -1$, recovering the marginal deformation mentioned in the main text.

Let us also consider continuous representation  and their spectral flow $\hat{\mathcal{C}}_j^{\alpha,w} \otimes \hat{\mathcal{C}}_j^{\alpha,w}$. For $s=\bar{s}=0$, the mass-shell condition is 
\begin{align}
\frac{1}{4k} + \frac{p^2}{k} + \frac{kw^2}{4} - n w + \frac{n^2}{4k} + N &= 1 \cr
\frac{1}{4k} + \frac{p^2}{k} + \frac{kw^2}{4} + nw + \frac{n^2}{4k} + \bar{N} &= 1 \ ,
\end{align}
where $j = \frac{1}{2} + ip$. Again we are interested in states which is not affected by the small change of $k$ (especially in the large $k$ limit), so we have to set $w=0$ to satisfy the mass-shell condition. Then for $n=0$, we have 
\begin{align}
|j = {1}/{2}+i\sqrt{k-{1}/{4}}, \alpha=0 \rangle \otimes |0 \rangle
\end{align}
and
\begin{align}
J^{+}_0 \tilde{J}^+_0 |j= {1}/{2}+ i\sqrt{{k}/{2}-{1}/{4}} , \alpha = 0 \rangle \otimes \psi_{-\frac{1}{2}} \bar{\tilde{\psi}}_{-\frac{1}{2}} |0 \rangle  \ .
\end{align}
Both of them do not preserve $\mathcal{N}=2$ supersymmetry (the latter series especially break $R$-symmetry).
Similar states exist for non-zero $n$ as $p=\sqrt{k-(1+n^2)/4k}$ or $p=\sqrt{k/2 - (1+n^2)/4k}$, example of which for $n=1$ is
\begin{eqnarray}
J^{-}_0 | j = 1/2 + i\sqrt{k-1/2}, \alpha = 1/2 \rangle \otimes |0\rangle \ .
\end{eqnarray}
However, neither of them preserve $\mathcal{N}=2$ supersymmetry, so they are not important for our studies.

\section{$\mathcal{N}=2$ Liouville theory}
We present our conventional form of $\mathcal{N}=2$ Liouville action by using $(2,2)$ superfield as (see \cite{Nakayama:2004vk} for details)
\begin{align}
S &= \frac{1}{4\pi}\int d^2 z d^4 \theta S \bar{S} + \mu_L \int d^2z d^2\theta e^{bS} + h.c. \cr
&= \frac{1}{2\pi} \int d^2z \partial \phi \bar{\partial}\phi + \partial Y \bar{\partial} Y + \bar{\psi}_+ \partial_- \psi_+ + \bar{\psi}_- \partial_+ \psi_- \cr
&+ \int d^2z \left(\mu b^2 \psi_+ \psi_- e^{b(\phi+iY)} + \bar{\mu} b^2 \bar{\psi}_+ \bar{\psi}_- e^{b(\phi-iY)} + \pi|\mu|^2 b^2 :e^{b(\phi+iY)}::e^{b(\phi-iY)}: \right) \ ,
\end{align}
where $S= \phi + iY + i\theta^+ \psi_- - i\theta^- \psi^+ +\cdots$ is a $(2,2)$ chiral superfield. The last term $ \pi|\mu|^2 b^2 :e^{b(\phi+iY)}::e^{b(\phi-iY)}:$ is a singular contact term and does not appear in the most of the CFT computation. There is a further linear dilaton coupling $\int d^2 z\sqrt{g} \mathcal{Q} \mathcal{R} \phi$ with $\mathcal{Q} = \frac{1}{b}$. The central charge is $c = 3 + \frac{3}{b^2}$, so the duality map is $b^2 = \frac{k}{2}$.

$(2,2)$ supersymmetry is generated by
\begin{align}
  T &= -\frac{1}{2}(\partial Y)^2 - \frac{1}{2}(\partial \phi)^2 + \frac{1}{2n}\partial^2 \phi -\frac{1}{4}(\psi_+\partial \bar{\psi}_+-\partial \psi_+ \bar{\psi}_+) \cr
G &= -\frac{1}{2}\psi_+(i\partial Y + \partial \phi) + \frac{1}{2b} \partial \psi_+ \cr
\bar{G} &=-\frac{1}{2}\bar{\psi}_+(i\partial Y - \partial \phi) - \frac{1}{2b} \partial \bar{\psi}_+ \cr
J &= \frac{1}{2}\psi_+ \bar{\psi}_+ + \frac{1}{b} i\partial Y
\end{align}
and similarly for the right-mover.


\begin{thebibliography}{99}
\bibitem{Distler:1987ee}
  J.~Distler and B.~R.~Greene,
  Nucl.\ Phys.\  B {\bf 304}, 1 (1988).


\bibitem{Beasley:2003fx}
  C.~Beasley and E.~Witten,
  JHEP {\bf 0310}, 065 (2003)
  [arXiv:hep-th/0304115].
\bibitem{Silverstein:1995re}
  E.~Silverstein and E.~Witten,
  Nucl.\ Phys.\  B {\bf 444}, 161 (1995)
  [arXiv:hep-th/9503212].

\bibitem{Basu:2003bq}
  A.~Basu and S.~Sethi,
  Phys.\ Rev.\  D {\bf 68}, 025003 (2003)
  [arXiv:hep-th/0303066].


\bibitem{Hori:2003ic}
  K.~Hori {\it et al.},
{\it  Providence, USA: AMS (2003) 929 p}

\bibitem{Blumenhagen:1996vu}
  R.~Blumenhagen, R.~Schimmrigk and A.~Wisskirchen,
  Nucl.\ Phys.\  B {\bf 486}, 598 (1997)
  [arXiv:hep-th/9609167].


\bibitem{Blumenhagen:1996tv}
  R.~Blumenhagen and S.~Sethi,
  Nucl.\ Phys.\  B {\bf 491}, 263 (1997)
  [arXiv:hep-th/9611172].

\bibitem{Blumenhagen:1997pp}
  R.~Blumenhagen and M.~Flohr,
  Phys.\ Lett.\  B {\bf 404}, 41 (1997)
  [arXiv:hep-th/9702199].
\bibitem{Knutson:1997yt}
  A.~Knutson and E.~R.~Sharpe,
  Adv.\ Theor.\ Math.\ Phys.\  {\bf 2}, 865 (1998)
  [arXiv:hep-th/9711036].
\bibitem{Sharpe:1998wa}
  E.~R.~Sharpe,
  arXiv:hep-th/9804066.

\bibitem{Adams:2003zy}
  A.~Adams, A.~Basu and S.~Sethi,
  Adv.\ Theor.\ Math.\ Phys.\  {\bf 7}, 865 (2004)
  [arXiv:hep-th/0309226].
\bibitem{Guffin:2007mp}
  J.~Guffin and S.~Katz,
  arXiv:0710.2354 [hep-th].

\bibitem{Melnikov:2007xi}
  I.~V.~Melnikov and S.~Sethi,
  JHEP {\bf 0803}, 040 (2008)
  [arXiv:0712.1058 [hep-th]].
\bibitem{Guffin:2008pi}
  J.~Guffin and E.~Sharpe,
  arXiv:0801.3955 [hep-th].
\bibitem{McOrist:2008ji}
  J.~McOrist and I.~Melnikov,
  arXiv:0810.0012 [hep-th].

\bibitem{FZZ}
V.~Fateev A.~Zamolodchikov, and A.~Zamolodchikov, unpublished.
\bibitem{Nakayama:2004vk}
  Y.~Nakayama,
  Int.\ J.\ Mod.\ Phys.\  A {\bf 19}, 2771 (2004)
  [arXiv:hep-th/0402009].
\bibitem{Hori:2001ax}
  K.~Hori and A.~Kapustin,
  JHEP {\bf 0108}, 045 (2001)
  [arXiv:hep-th/0104202].

\bibitem{Witten:1993yc}
  E.~Witten,
  Nucl.\ Phys.\  B {\bf 403}, 159 (1993)
  [arXiv:hep-th/9301042].

\bibitem{Hori:2000kt}
  K.~Hori and C.~Vafa,
  arXiv:hep-th/0002222.

\bibitem{2DBH}
G S. Elitzur, A. Forge and E. Rabinovici,
Nucl. Phys. {\bf B359} (1991) 581;
Mandal, A. Sengupta and S. Wadia,
Mod. Phys. Lett. {\bf A6} (1991) 1685;
E. Witten,
Phys. Rev. {\bf D44} (1991) 314.

\bibitem{Nakayama:2006gt}
  Y.~Nakayama, S.~J.~Rey and Y.~Sugawara,
  arXiv:hep-th/0606127.


\bibitem{Ghoshal:1995wm}
  D.~Ghoshal and C.~Vafa,
  Nucl.\ Phys.\  B {\bf 453}, 121 (1995)
  [arXiv:hep-th/9506122].
\bibitem{Giveon:1999zm}
  A.~Giveon, D.~Kutasov and O.~Pelc,
  JHEP {\bf 9910}, 035 (1999)
  [arXiv:hep-th/9907178].

\bibitem{Lerche:2000uy}
  W.~Lerche,
  arXiv:hep-th/0006100.
\bibitem{Eguchi:2004ik}
  T.~Eguchi and Y.~Sugawara,
  JHEP {\bf 0501}, 027 (2005)
  [arXiv:hep-th/0411041].
\bibitem{Hori:2002cd}
  K.~Hori and A.~Kapustin,
  JHEP {\bf 0211}, 038 (2002)
  [arXiv:hep-th/0203147].
\bibitem{Nakayama:2004at}
  Y.~Nakayama,
  Nucl.\ Phys.\  B {\bf 708}, 345 (2005)
  [arXiv:hep-th/0409039].

\bibitem{Eguchi:2004yi}
  T.~Eguchi and Y.~Sugawara,
  JHEP {\bf 0405}, 014 (2004)
  [arXiv:hep-th/0403193].

\bibitem{Israel:2004ir}
  D.~Israel, C.~Kounnas, A.~Pakman and J.~Troost,
  JHEP {\bf 0406}, 033 (2004)
  [arXiv:hep-th/0403237].
\bibitem{Kazama:1988qp}
  Y.~Kazama and H.~Suzuki,
  Nucl.\ Phys.\  B {\bf 321}, 232 (1989).
\bibitem{Kazama:1988uz}
  Y.~Kazama and H.~Suzuki,
  Phys.\ Lett.\  B {\bf 216}, 112 (1989).

\bibitem{Maldacena:2000hw}
  J.~M.~Maldacena and H.~Ooguri,
  J.\ Math.\ Phys.\  {\bf 42}, 2929 (2001)
  [arXiv:hep-th/0001053].

\end{thebibliography}
\end{document}